\documentclass[showpacs,aps,twocolumn]{revtex4}
\usepackage{epsfig}
\usepackage{graphicx}
\usepackage{amsmath,amssymb,amsfonts}
\usepackage{array}
\usepackage{url}
\usepackage{hyperref}
\usepackage{multirow}
\usepackage{float}
\usepackage{lineno}
\usepackage{xspace}
\usepackage[usenames,dvipsnames]{color}
\usepackage{adjustbox}

\newcommand{\bef}{\begin{ure}}
\newcommand{\eef}{\end{ure}}
\newcommand{\bc}{\begin{center}}
\newcommand{\ec}{\end{center}}

\newcommand{\be}{\begin{equation}}
\newcommand{\ee}{\end{equation}}
\newcommand{\bea}{\begin{eqnarray}}
\newcommand{\eea}{\end{eqnarray}}

\def\ba{\begin{eqnarray}}
\def\ea{\end{eqnarray}}
\begin{document}

\title{Role of chemical potential at kinetic freeze-out using Tsallis non-extensive statistics in proton-proton collisions at the Large Hadron Collider}

\author{Girija Sankar Pradhan$^{1}$}
\author{Dushmanta Sahu$^{1}$}
\author{Rutuparna Rath$^{2}$}
\author{Raghunath Sahoo$^{1}$\footnote{Corresponding Author Email: Raghunath.Sahoo@cern.ch}}
\author{Jean Cleymans$^{3}$\footnote{Deceased}}
\affiliation{Department of Physics, Indian Institute of Technology Indore, Simrol, Indore 453552, India}
\affiliation{$^{2}$INFN - sezione di Bologna, via Irnerio 46, 40126 Bologna BO, Italy}
\affiliation{$^{3}$UCT-CERN Research Centre and Physics Department, University of Cape Town, Rondebosch 7701, South Africa}

\begin{abstract}
The charged-particle transverse momentum spectra ($p_{\rm T}$-spectra) measured by the ALICE collaboration for $pp$ collisions at $\sqrt {s} =$ 7 and 13 TeV have been studied using a thermodynamically consistent form of Tsallis non-extensive statistics. The Tsallis distribution function is fitted to the $p_{\rm T}$-spectra and the results are analyzed as a function of final state charged-particle multiplicity for various light flavor and strange particles, such as $\pi^{\pm}, K^{\pm}, p+\bar{p}, \phi, \Lambda+\bar{\Lambda}, \Xi+\bar{\Xi}, \Omega+\bar{\Omega}$. At the LHC energies, particles and antiparticles are produced in equal numbers. However, the equality of particle and antiparticle yields at the kinetic freeze-out may imply that they have the same but opposite chemical potential which is not necessarily zero. We use an alternative procedure that makes use of parameter redundancy, by introducing a finite chemical potential at the kinetic freeze-out stage. This article emphasizes the importance of the chemical potential of the system produced in $pp$ collisions at the LHC energies using the Tsallis distribution function which brings the system to a single freeze-out scenario.

\end{abstract}
\date{\today}
\maketitle{}

\section{Introduction}
\label{intro}

It has been the ambitious force of research in high-energy physics to use particle colliders like the Relativistic Heavy Ion Collider (RHIC) at Brookhaven National Laboratory (BNL) and the Large Hadron Collider (LHC) at CERN (European Council for Nuclear Research), which fulfill the appetite for understanding the matter formed during ultra-relativistic collisions. The primary objective of these experiments is to create extreme conditions of high temperature and/or energy densities through compression or heating in high-energy nuclear collisions~\cite{Gyulassy:2004zy, Braun-Munzinger:2007edi, Jacak:2012dx}, where a system of deconfined quarks and gluons can be formed. These extreme conditions lead to the asymptotic freedom of the QCD system, where the quarks and gluons are no longer confined inside the nucleon \cite{Itoh:1970uw, Collins:1974ky, Cabibbo:1975ig, Chapline:1976gy}. Afterward, the produced system undergoes expansion through cooling of the systems where the quarks and gluons are combined to form hadrons. In this course of action, the inelastic collisions cease at the chemical freeze-out boundary where the stable particle yields get fixed at chemical freeze-out temperature ($T_{\rm ch}$). Finally, at the kinetic freeze-out boundary, the elastic collisions among the final state particles no longer exist, and a stream of particles gets detected in the detectors. This enables us to get the kinetic freeze-out temperature ($T$) of the system from the transverse momentum spectra of the identified particles. In literature, various studies have been performed to estimate the temperature at different stages of evolution, which includes the initial temperature ($T_{\rm i}$)~\cite{Waqas:2019bnc,Wang:2021ccs}, chemical freeze-out temperature ($T_{\rm ch}$)~\cite{Cleymans:1999st,Florkowski:1999pz,Braun-Munzinger:2003htr}, and kinetic freeze-out temperature ($T$)~\cite{Waqas:2018xrz,Waqas:2021bsy,Waqas:2021enm,Ajaz:2021awb}. The chemical freeze-out stage is well understood and is strongly supported by experimental results~\cite{Andronic:2017pug}. Furthermore, the results are in agreement with the Lattice Quantum Chromodynamics (LQCD) model based on first principles and a well-established hydrodynamic technique~\cite{Sollfrank:1990qz}.

Similarly, the kinetic freeze-out stage has also been explored by several phenomenological works, where information about the system is extracted by fitting different functions to the transverse momentum spectra of the final state particles. The kinetic freeze-out of particles is a highly complex phenomenon. For instance, Ref.~\cite{Tang:2008ud} shows the single kinetic freeze-out scenario, whereas Ref.~\cite{Chatterjee:2014lfa,Chatterjee:2015fua,Thakur:2016boy} shows a double kinetic freeze-out scenario, and Ref.~\cite{Chatterjee:2014ysa,Waqas:2018tkk,Waqas:2021jph} infers about the multiple kinetic freeze-out scenarios. Furthermore, a recent study~\cite{Waqas:2021rmb} observed a triple kinetic freeze-out scenario for nonstrange, strange, and multi-strange particles and found that the multi-strange particles freeze out earlier than the strange and nonstrange particles. Although studies on the kinetic freeze-out scenarios have concluded differently, it is still an open question in the high-energy physics community. Moreover, studies show that finite radial flow~\cite{ALICE:2020nkc} and collective effects such as long-range correlations~\cite{CMS:2015fgy} have been observed in high-energy pp collisions. These phenomena are often attributed to the underlying partonic structure of the colliding protons rather than radial flow. Thus, this article shows an alternative procedure to the single freeze-out scenario without taking the radial flow effect. Here, we present an alternative approach where the flow effects are neglected. 

Over the past several years, numerous studies of lattice QCD have been successful at high temperatures and with vanishing chemical potential. However, studying the phase structure of QCD at non-zero chemical potential is one of the most exciting problems in contemporary physics~\cite{Allton:2002zi, Allton:2003vx, Gavai:2003mf, Allton:2005gk, Bazavov:2017dus}. It is noteworthy that on the theoretical side, the color superconducting and superfluid phases have been conjectured at high baryon densities~\cite{Book}. Therefore, it is necessary to investigate the QCD phase transition utilizing lattice gauge theory simulations at the non-zero chemical potential. As discussed in Ref. ~\cite{Andronic:2017pug}, a consequence of the vanishing baryon-chemical potential leads to the vanishing of strangeness chemical potential {$\mu_s$}, which implies that the strange quantum number is no longer relevant for particle production. The yield of strange and multi-strange mesons and baryons in the fireball is solely determined by their mass, $m$, spin degeneracy, $g$, and the temperature, $T$. At the LHC energies near chemical freeze-out, the baryochemical potential is expected to be zero due to particles and antiparticles being produced in equal numbers. However, assuming the same case at the kinetic freeze-out temperature is not entirely trivial. Thus, there can be a finite total chemical potential at the kinetic freeze-out stage, and its implications cannot be ignored. Specifically, this study considers the importance of chemical potentials at the kinetic freeze-out stage.

As discussed earlier, the $p_{\rm T}$-spectra measured in $pp$ collisions at the LHC energies can give information about the kinetic freeze-out stage of the collision. We generally fit a Boltzmann-type distribution function to the $p_{\rm T}$-spectra to extract useful information. However, the Boltzmann distribution function can only explain the low-$p_{\rm T}$ part of the spectra. As we know, transverse momenta up to hundreds of GeV have been measured in high-energy hadronic collisions. This suggests that the high-$p_{\rm T}$ regime is also very significant to understand the system formed in such collisions and cannot be ignored. Power-law type distribution functions can describe the high-$p_{\rm T}$ part of the spectra very well, which comes from the perturbative QCD. However, a unique distribution function is needed to understand the system thoroughly, which can explain both the low- and high-$p_{\rm T}$ part of the spectra. The thermodynamically consistent Tsallis distribution function has been widely used for this purpose~\cite{Abelev:2006cs, PHENIX:2010qqf, PHENIX:2011rvu, CMS:2010wcx, CMS:2010tjh, ATLAS:2010jvh, Aamodt:2011zj, Abelev:2012cn, Abelev:2012jp, Chatrchyan:2012qb}. The thermodynamic consistency of $q$-exponent Tsallis distributions can be found in Ref.~\cite{Cleymans:2011in,Cleymans:2012ya,Azmi:2015xqa,Pradhan:2021vtp}. In this view, the formulation of the non-extensive $q$-generalized statistics must be based on the non-additive Tsallis entropy~\cite{Tsallis:1998ws,Tsallis:1987eu,Lyra:1997ggy,Tsallis:2012js,Luciano:2021ndh,Tsallis:book}. It is worth noting that the $q$-generalized Tsallis statistics has provided encouraging results in describing a broad class of complex systems, such as self-gravitating stellar systems~\cite{Plastino}, black holes~\cite{Tsallis:2012js}, cosmic background radiation~\cite{Tsallis:1995zza}, low-dimensional dissipative systems~\cite{Lyra:1997ggy}, and polymer chains~\cite{Jizba}, among others. An adequate description of field mixing based on Tsallis entropy has recently been proposed in~\cite{Luciano:2021mto}. However, the only drawback of this distribution is not being Lorentz invariant. It has been found that the expectation values are not consistent with the normalization condition of probabilities and are not invariant under the overall shift in energy. Please refer to Ref.~\cite{Parvan:2021rhi, Parvan:2019aii} for detailed calculations and explanations. It should be noted here that there are various forms of Tsallis distribution function
as discussed in Ref. \cite{Parvan:2021rhi,Parvan:2019aii}. However, our choice of the function is motivated by the
thermodynamic consistency of the function, as discussed in the paper.

Furthermore, the multiplicity fluctuations in high-energy nuclear collisions attribute them to intrinsic fluctuations of the temperature of the hadronizing system formed in such processes. To account for these fluctuations,  the non-extensivity parameter $q$ is used, where $|q-1|$ is a direct measure of fluctuation \cite{Wilk:1999dr,Biro:2004qg,Wilk:2009nn,Wilk:2008ue,Biro:2014yoa}. Here the non-extensive parameter $q$ is related to the event-by-event temperature fluctuation of a system, and is given by $ q = 1 + Var(1/T)/\langle 1/T\rangle ^2 $. To study the evolution of temperature fluctuation in a system approaching thermodynamic equilibrium, the Boltzmann transport equation in Tsallis non-extensive statistics is also used \cite{Bhattacharyya:2015nwa}. As discussed in Ref. \cite{Bhattacharyya:2015nwa}, the QCD deconfinement transition and the location of the possible critical point in the QCD phase diagram are associated with large-scale observable fluctuations, which should be reflected in the fluctuations in $q$-values. In addition, the non-extensivity of a physical system produced in high-energy hadronic and heavy-ion collisions
is also linked to temperature fluctuation, and hence, the heat capacity/specific heat of the system.

Moreover, the connections between Boltzmann and Tsallis statistics proposed so far are related to thermodynamical aspects of the system but not directly related to the microscopic aspects of hadronic matter and QCD interaction. A comparison of results from the non-extensive self-consistent thermodynamics LQCD has been performed \cite{Deppman:2012qt,Deppman:2016fxs,Deppman:2019yno,Deppman:2017fkq} showing a fair agreement between the two methods. A recent investigation \cite{Deppman:2016fxs} introduces a system with a fractal structure in its thermodynamical functions called thermofractal. It is shown that its thermodynamics is more naturally described by Tsallis than the Boltzmann statistics. A relation between the fractal dimension and the entropic index ($q$) has been established. The ratio between the Tsallis temperature ($\tau$) and the Boltzmann temperature ($T$) is linked to the entropic index ($q$) and the number of subsystems. It is shown that while $\tau$ regulates the system energy, $T$ regulates the fraction of the total energy accumulated as the internal energy of the subsystems. On the one hand, this result allows one to connect the entropic index to fundamental aspects of the interaction between the constituents and, on the other hand, to establish constraints on the S-matrix to allow the emergence of non-extensivity in the corresponding system.

The non-extensive parameter, $q$, in the Tsallis distribution function quantifies the degree of deviation of the system from the equilibrium state. In addition, Tsallis temperature ($T$), volume ($V$), and the chemical potential ($\mu$) can also be extracted by fitting the Tsallis distribution to the transverse momentum spectra. 

This article is organized as follows. In the next section~\ref{formulation}, we explicitly discuss the single-particle Tsallis distribution to fit the transverse momentum spectra of the identified hadrons and determine the temperature $T$, volume $V$, and $q$ with the details of the formulation of estimating the chemical potential from kinetic freeze-out. In section~\ref{res}, we discuss our results, and finally, we summarize our results and conclude in section~\ref{sum}.     

\section{Formulation}
\label{formulation}
The Tsallis distribution function that satisfies the thermodynamic consistency relations~\cite{Cleymans:2011in,Cleymans:2012ya,Azmi:2015xqa} is given by,

\begin{equation}
\label{InvariantYield}
E\frac{d^3N}{dp^3} = gV E\frac{1}{(2 \pi)^3} \left[ 1 + ( q - 1) \frac{ E - \mu}{T}\right]^ {-\frac{q}{q -1}}.
\end{equation}

Where, $E$ represents particles' energy, $d^3N$ is the invariant yield, $V$ is the volume of the system, $g$ is the degeneracy factor, $q$ is referred to as the non-extensive parameter, $T$ is the corresponding temperature, $p$ denotes the momentum, and $\mu$ is the total chemical potential 
($\mu = B\mu_B + S\mu_S + Q\mu_Q$). Here $B$ is the baryon number, $S$ is the strangeness quantum number, and $Q$ is the electric charge. $\mu_B$, $\mu_S$, and $\mu_Q$ represent the baryon, strangeness, and electric charge chemical potential, respectively. Physical interpretation of $q$ can be found in Ref.~\cite{Wilk:1999dr}. At mid rapidity, {\it i.e,} {$y = 0$}, Eqn.~\ref{InvariantYield} in terms of transverse momentum, $p_{\rm T}$, transverse mass, $ m_T = \sqrt{p_T^2 + m^2}$, where $E = m_{T} coshy$ can be rewritten as, 

\begin{equation}
\label{YieldNonZeroMu}
\frac{d^2N}{dp_Tdy}\bigg|_{y = 0}= gV \frac{p_Tm_T }{(2 \pi)^2} \left[ 1 + ( q - 1) \frac{m_T - \mu}{T}\right]^ {-\frac{q}{q -1}}.
\end{equation}

In order to extract Tsallis parameters for the identified non-strange, strange, and multi-strange particles, Eqn.~\ref{YieldNonZeroMu} has been used. The degeneracy factor $g = 2 \times(2s + 1)$ is taken to be 2, 2, 4, 3, 8, 4 and 8 for $\pi^{\pm}, K^{\pm}, p+\bar{p}, \phi, \Lambda+\bar{\Lambda}, \Xi+\bar{\Xi}, \Omega+\bar{\Omega}$, respectively. Here, $s$ is the particle's spin, and factor 2 is for the antiparticles. $\Sigma^{0}$ and $\Lambda$ are not experimentally distinguishable. Thus, the degeneracy factor for the $\Lambda$ particle is 8. It is worth noting that the four parameters $T,V,q$ and $\mu$ in Eqn.~\ref{YieldNonZeroMu} has a redundancy for $\mu \neq 0$ in Ref.~\cite{Cleymans:2013rfq,Cleymans:2020ojr,Cleymans:2020nvs,Rybczynski:2014cha}. Precisely for a fixed values of $q$, let $T = T_{0}$ and $V = V_{0}$ at $\mu = 0$. So, comparing Eqn.~\ref{YieldNonZeroMu} for $\mu = 0$ and for a finite value of $\mu$, we obtain the following transformation relations~\cite{Cleymans:2013rfq,Cleymans:2020ojr,Cleymans:2020nvs,Rybczynski:2014cha}.

\begin{eqnarray}
T_0 &=& T \left[1-(q-1) \frac{\mu}{T} \right], \text{with} \qquad \mu\leq\frac{T}{q-1},
\label{T0} \\
V_0 &=& V \left[1-(q-1) \frac{\mu}{T} \right]^{\frac{q}{1-q}}.
\label{V0}
\end{eqnarray}

Hence, the variables $T$ and $V$ are functions of $\mu$ at fixed values of $q$ and can be determined if the parameters ($T_{0}$, $V_{0}$) and $q$ are known. This redundancy is not present when $\mu = 0$. Then, the transverse momentum distribution in terms of these modified variables can be written as,

\begin{equation}
\label{YieldIndexZero}
\frac{d^2N}{dp_Tdy} = gV_0 \frac{p_Tm_T}{(2 \pi)^2} \left[ 1 + ( q - 1) \frac{m_T}{T_0}\right]^ {-\frac{q}{q -1}}
\end{equation}

Here, the system's chemical potential ($\mu$) does not appear explicitly. Analogous to the volumes $V$ and $V_0$ defined in Eqn.~\ref{InvariantYield} and~\ref{V0}, we also introduce the corresponding radii $R$, and $R_0$, assuming a spherically symmetric system~\cite{Okorokov:2014cna, Okorokov:2016vug}.

\begin{eqnarray}
V &=& \frac{4\pi}{3}R^3 ,
\label{R}\\
V_0 &=& \frac{4\pi}{3}R_0^3.
\label{R0}
\end{eqnarray}

The two-particle interferometry analysis (often referred to as HBT), based on Bose-Einstein correlations, is a unique experimental method for determining the sizes and lifetime of particle sources in high energy and nuclear physics. The value of $R$ is not necessarily related to the size of the system as deduced from an HBT analysis~\cite{Okorokov:2014cna, Okorokov:2016vug,ALICE:2010igk, STAR:2010yvd}; however, it serves to fix the normalization of the distribution~\ref{YieldNonZeroMu}. An important suggestion was to determine the chemical potential in~\cite{Rybczynski:2014cha}, where the observation was that the radius $R_0$ given in Table~ IV and V is larger than one obtained from a femtoscopy analysis~\cite{ALICE:2011kmy} by a factor $\kappa$ estimated to be about 3.5, i.e.,

\begin{equation}
R_{\rm Femto} \approx \frac{1}{\kappa}R_0.
\label{rFemto}
\end{equation}

Hence, in~\cite{Rybczynski:2014cha}, a suggestion is made to identify the corresponding volume $V_{\rm Femto}$ with the volume $V$ appearing in Eqn.~\ref{InvariantYield}.

Hence
\begin{equation}
\label{rFemto_Assumption}
V_0 \approx V\cdot \kappa^3.
\end{equation}

Combining this with Eqn.~$\ref{T0}$ and $\ref{V0}$ leads to a chemical potential given by
\begin{equation}
\label{Assumption_mu}
\mu = \frac{T_0}{q -1} \left( \kappa^{3(q -1)/q} - 1 \right),
\end{equation}

Hence, using this proposal, a knowledge of $T_0$ would determine $\mu$. Please refer to Ref.~\cite{Rybczynski:2014cha} for a more detailed discussion. 

In this work, we compared the values of the chemical potential $\mu$ using this proposal~\cite{Rybczynski:2014cha} to the values using the procedure outlined above starting Eqn.~\ref{YieldNonZeroMu} and concluded that the results are very different. Thus, our results do not support this assumption, and therefore, the volume $V$ appearing in Eqn.~\ref{YieldNonZeroMu} cannot be identified with the volume determined from femtoscopy. $V$ must be considered specific to the Tsallis distribution, as with all the other variables used in this paper.

At chemical equilibrium, one has $\mu = 0$ for all quantum numbers as the number of particles and antiparticles are equal. However, the equality of particle-antiparticle at kinetic freeze-out implies equal chemical potential but not necessarily zero. We stress that Eqns.~\ref{YieldNonZeroMu} and~\ref{YieldIndexZero} carry different meanings where neither $T_0$ is not equal to $T$, nor is $V_0$ equal to $V$. It is worth noting that we do not have $\mu$ in Eqn.~\ref{YieldIndexZero}.

The purpose of the current paper is to resolve this issue. For this, we choose the following technique:

\begin{enumerate}
\item To determine the three parameters $T_0$, $q$, and $V_0$, we use Eqn.~\ref{YieldIndexZero} to fit the transverse momentum distribution, keeping all the parameters free.

\item Fix the value of parameter $q$, which is obtained from the previous step.

\item Then perform the fit to the transverse momentum distributions using Eqn.~\ref{YieldNonZeroMu}, keeping $q$ fixed as determined in the previous step, which determines the parameters $T$, $V$ and the chemical potential $\mu$.

\item We show that the choice of $q$, which is particle species dependent, appears to be independent of the chemical potential
of the system for all particles.

\item At last, check the consistency with Eqns.~\ref{T0} and $~\ref{V0}$.
\end{enumerate}

Each step of the fitting procedure includes only three parameters to describe the transverse momentum distributions. This method was presented in ~\cite{Cleymans:2020ojr,Cleymans:2020nvs}. The present work conveys that the chemical potential at kinetic freeze-out is not identical to that at chemical freeze-out. The chemical potentials are considered zero at chemical freeze-out, where thermal and chemical equilibrium has been established. At kinetic freeze-out, we observed a finite chemical potential for both particle and antiparticle. However, they do not have to be zero due to the absence of chemical equilibrium at kinetic freeze-out. The only limitation is that they should be equal for particles and antiparticles.

\section{Results and Discussion}

\label{res}
The $p_{\rm T}$-spectra for non-strange, strange, and multi-strange particles  $\pi^{\pm}, K^{\pm}, p+\bar{p}, \phi, \Lambda+\bar{\Lambda}, \Xi+\bar{\Xi},$ and $\Omega+\bar{\Omega}$ are fitted up to $p_{\rm T}$ = 6 GeV/c with a thermodynamically consistent form of Tsallis distribution function for $pp$ collisions at $\sqrt{s}$ = 7 TeV and $\sqrt{s}$ = 13 TeV for different multiplicity classes. The fitting is performed utilizing the TMinuit class available in the ROOT library, keeping all three parameters free \cite{ROOT}. We have used a chi-squared fitting procedure in this work. In the foremost step, we determine non-extensive parameter ($q$), temperature parameter ($T_{0}$) and radius parameter ($R_{0}$) at zero chemical potential ($\mu = 0$). In the subsequent step, we fix the non-extensive parameter, which is obtained from the first step, to extract all the fitting parameters, such as temperature ($T$), the radius of the system ($R$), and chemical potential ($\mu$). The quality of the fits is indicated by the reduced-$\chi^2$ listed in the tables. This indicates that the non-extensive distribution function well describes the spectra. However, the reduced-$\chi^2$ values are comparatively bad for pions in contrast to other particle species considered in this analysis, possibly due to the contribution from resonance decay.

\begin{figure*}[ht!]
\begin{center}
\includegraphics[scale = 0.44]{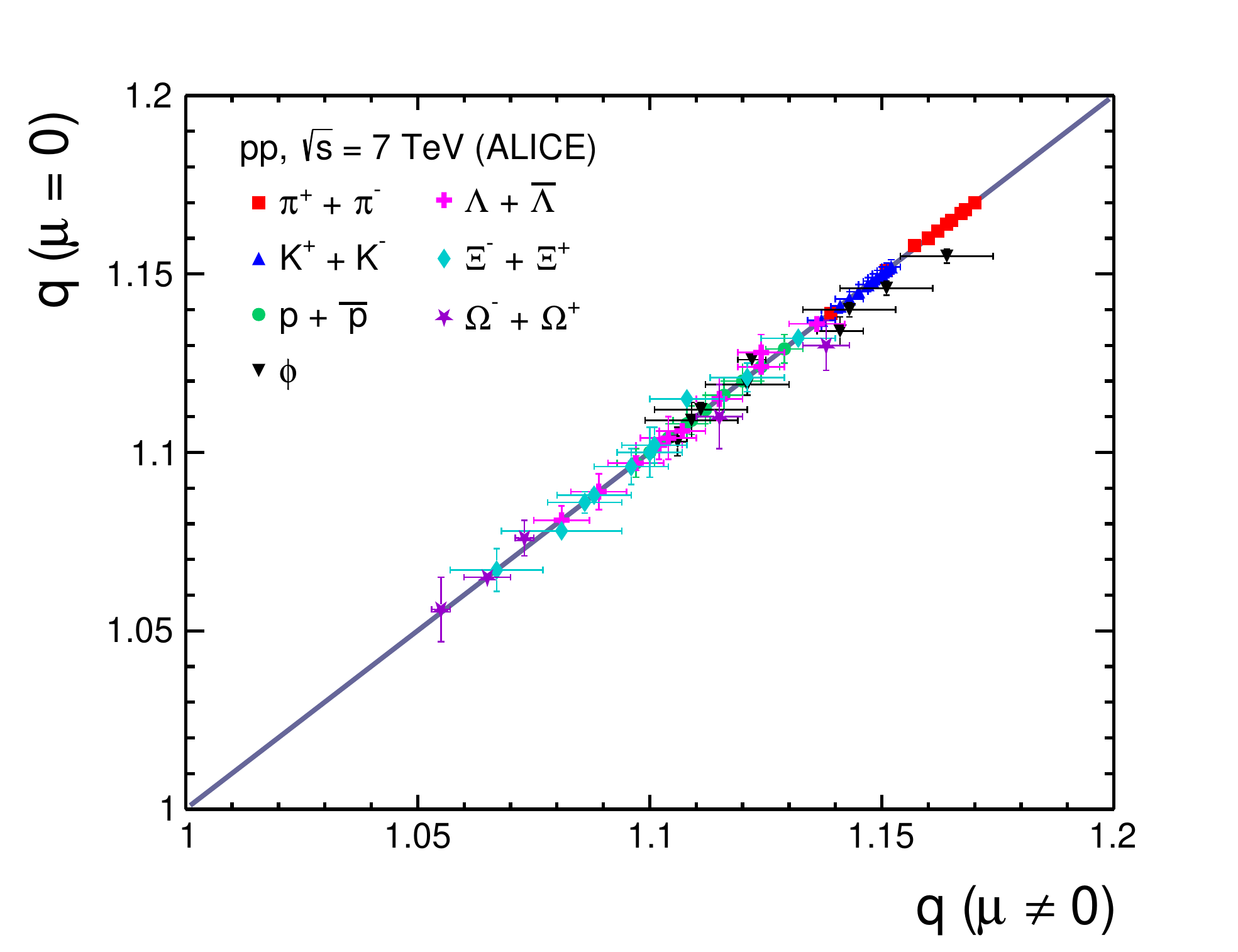}
\includegraphics[scale = 0.44]{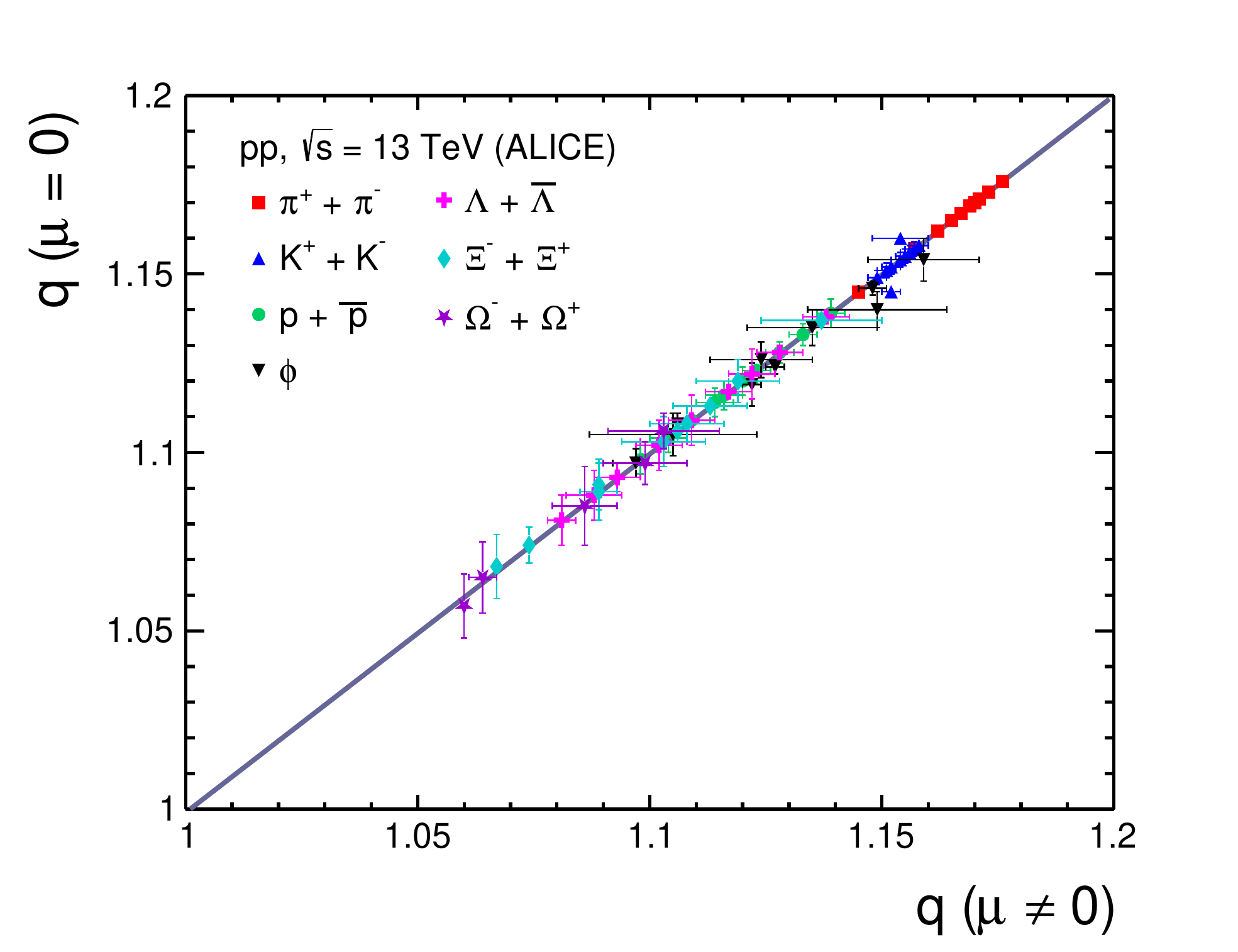}
\caption{(Color online) The non-extensive parameter ($q$) at ($\mu = 0$) as a function of $q$ for ($\mu \neq 0$) for $pp$ collisions at $\sqrt{s}$ = 7  TeV (left panel) and 13 TeV (right panel) for different final state particles. The line represents the fit function, $y = mx+c$.}
\label{fig1}
\end{center}
\end{figure*}

\begin{figure}[ht!]
\begin{center}
\includegraphics[scale = 0.45]{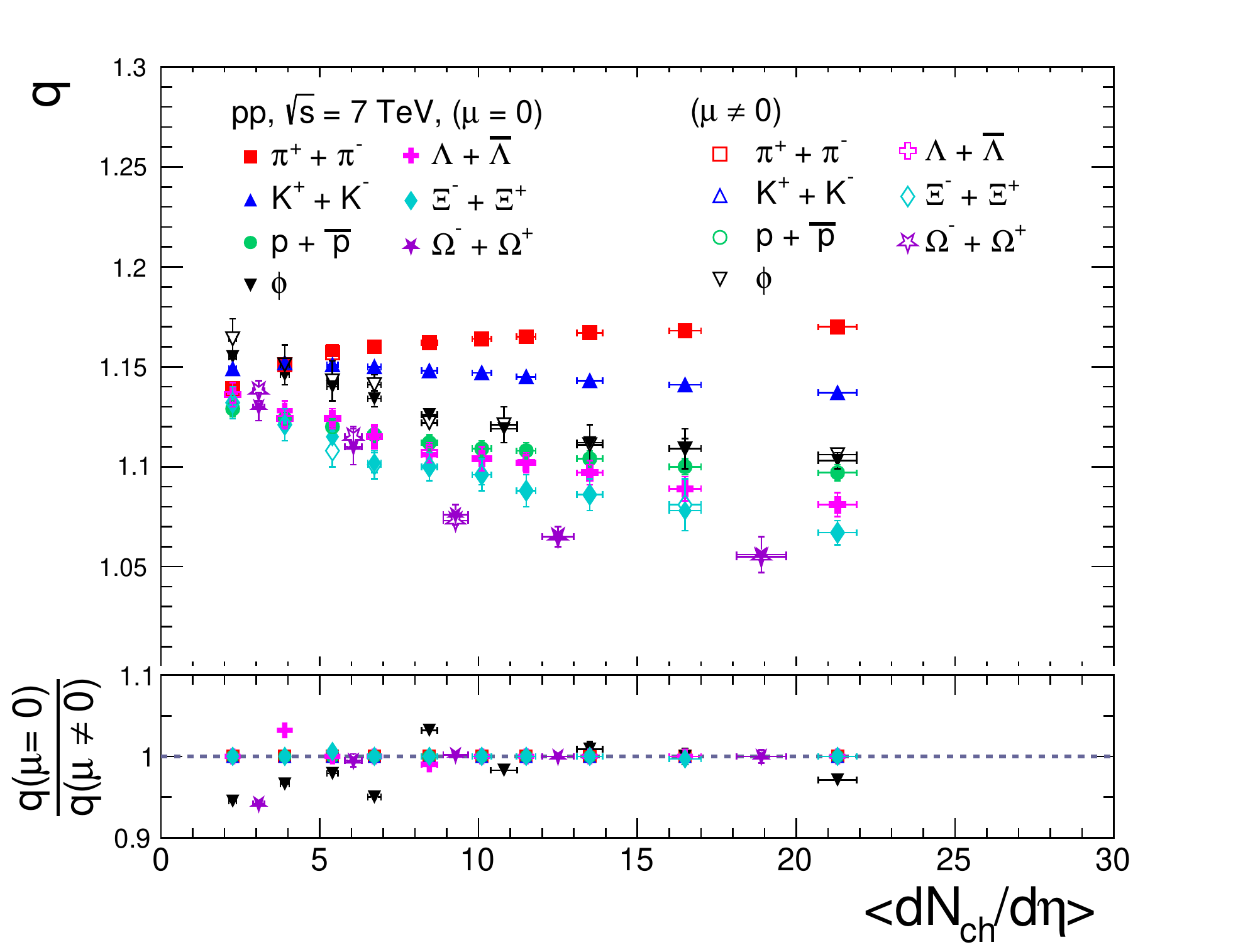}
\caption{(Color online) Comparison of the non-extensive parameter ($q$) at ($\mu = 0$) and ($\mu \neq 0$) for $pp$ collisions at $\sqrt{s}$ = 7  TeV for different final state particles as a function of final state multiplicity. Shown in the bottom panel is the ratio of both cases, which indicates that $q$ hardly depends on the chemical potential of the system.}
\label{fig2}
\end{center}
\end{figure}

Figure~\ref{fig1} shows the variation of the non-extensive parameter at ($\mu = 0$) and ($\mu \neq 0$) for $pp$ collisions at $\sqrt{s}$ = 7 TeV (left panel) and 13 TeV (right panel) for different final state particles. As in Section \ref{formulation} of the article, we have mentioned that the value of the non-extensive parameter $q$ is kept fixed, and we perform the fit to the transverse momentum distributions to determine the parameters like $T$, $V$, and the chemical potential $\mu$. So, to validate our method, we have plotted the value of $q$ for both the cases $\it{i.e.}$ $(\mu = 0$) and ($\mu \neq 0$). This suggests the value of $q$ is independent irrespective of the value of $\mu$. The contribution of $\mu$ is taken care of by $T_0$ and $V_0$ as mentioned in Eqns. \ref{T0} and \ref{V0}. We fitted the spectrum with $y = mx+c$ for both the centre-of-mass energies  $\it{i.e}$ $\sqrt{s}$ = 7 and 13 TeV.  The values of the parameters are: slope, $m = 0.998 \pm 0.019$ and the intercept, $c = 0.002 \pm 0.022$ for $\sqrt{s}$ = 7 TeV.  Similarly, $m = 1.004 \pm 0.018$ and $c = -0.005 \pm 0.021$ for $\sqrt{s}$ = 13 TeV.
Going one step further, we plot the variation of the $q$-parameter for both the cases of $\mu = 0$ and $\mu \neq 0$ for all the considered particle species as a function of the final state-charged particle multiplicity for $\sqrt{s}$ = 7 TeV. This is shown in Fig. \ref{fig2}. The bottom panel of which shows a ratio indicating a near-independency of the $q$-parameter on the chemical potential of the system. This justifies the procedure mentioned in the above section.

\begin{figure*}[ht!]
\begin{center}
\includegraphics[scale = 0.44]{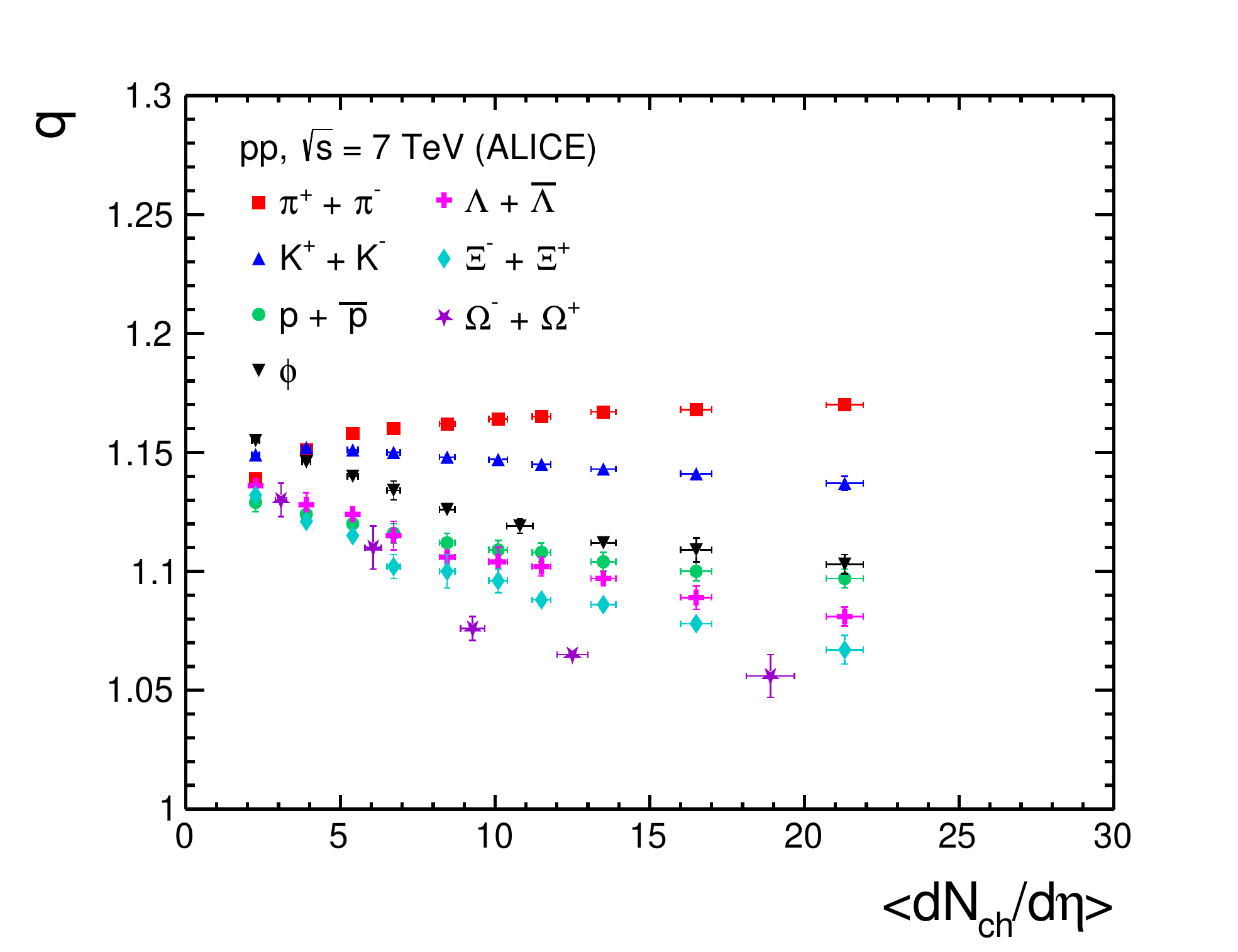}
\includegraphics[scale = 0.44]{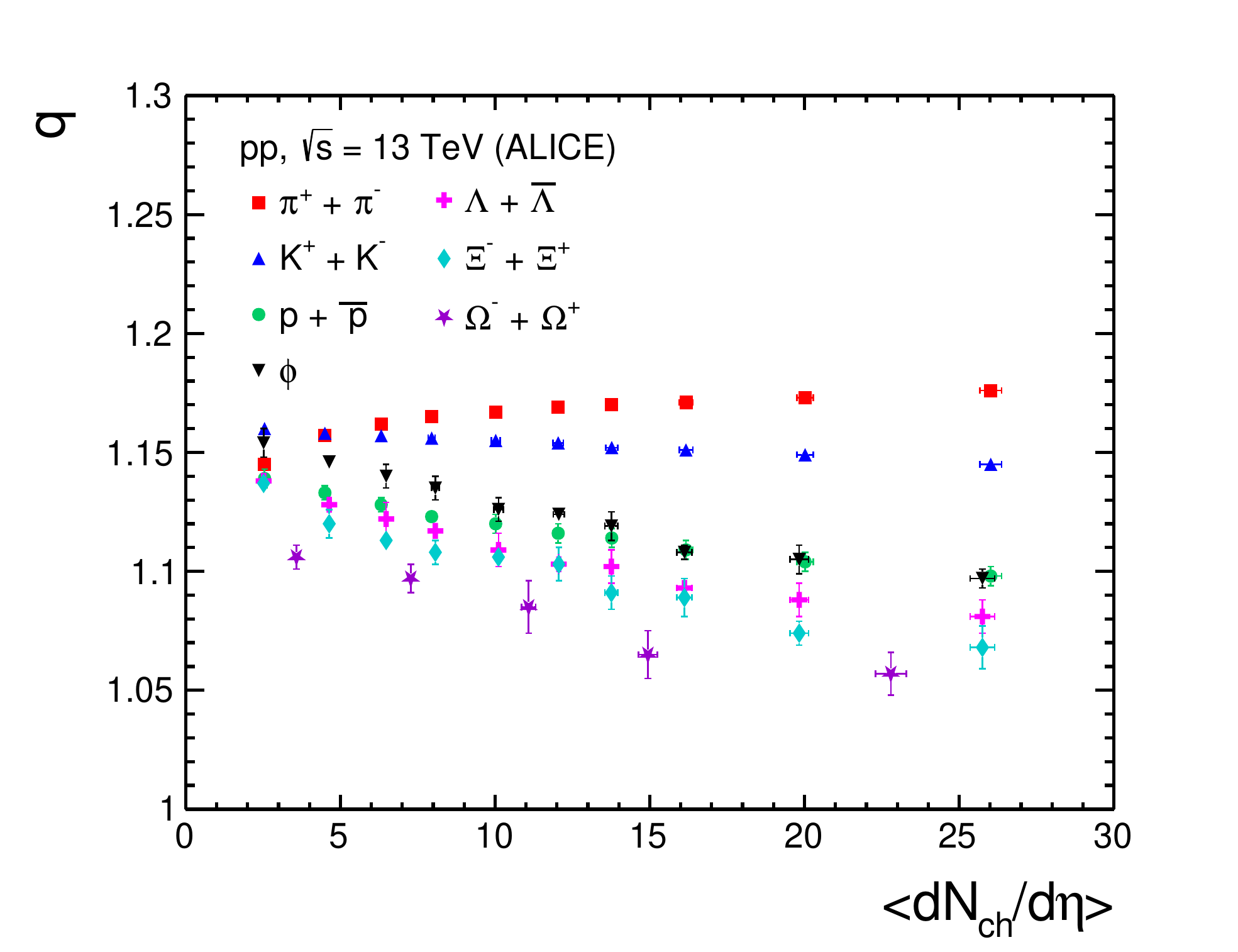}
\caption{(Color online) Non-extensive parameter ($q$) as a function of charged-particle multiplicity for $pp$ collisions at $\sqrt{s}$ = 7 TeV (left panel) and 13 TeV (right panel) for different final state particles. The uncertainties in charged-particle multiplicity are the quadratic sum of statistical and systematic contributions and the error in the value of $q$ are statistical errors.}
\label{fig3}
\end{center}
\end{figure*}

In Fig.~\ref{fig3}, we use Eqn.~\ref{YieldIndexZero} to fit the transverse momentum spectra of various identified particles observed experimentally for different multiplicity classes at $\sqrt{s}$ = 7 TeV and $\sqrt{s}$ = 13 TeV. It illustrates the variation of non-extensive parameter ($q$) as a function of charged-particle multiplicity at $\sqrt{s}$ = 7 (left panel) and 13 TeV (right panel) for various final state particles. The value of $q$ decreases monotonically with an increase in multiplicity for all the particles except for pions, suggesting that the system created in higher multiplicity classes is close to thermal equilibrium. It seems that the more important thing is $q$ approaches 1 for high multiplicities. The decrease in $q$-values is an important observation as it infers that the hot and dense system formed in high multiplicities becomes close to a thermalized Boltzmann description of the system. However, it can be observed that the value of $q$ monotonically increases for pions with charged-particle multiplicity. This suggests that the system deviates further from thermal equilibrium as the multiplicity of charged particles increases. It could be due to the contribution of resonance decay. Higher multiplicity might involve a more significant number of resonances, which can introduce additional non-equilibrium effects and influence the value of $q$.

 
 \begin{table*}[htbp]
\caption[p]{Number of mean charged particle multiplicity density for all the particles at mid rapidity corresponding to different event classes at $\sqrt{s}$ = 7 TeV~\cite{ALICE:2018pal}. The uncertainties are the quadratic sum of statistical and systematic contributions.} 
\label{table:mult_info}
\begin{adjustbox}{max width=\textwidth}
\begin{tabular}{c|c|c|c|c|c|c|c|c|c|c|c|}
\hline
\multicolumn{2}{|c|}{${\bf Multiplicity class}$}&Mul1&Mul2&Mul3&Mul4&Mul5&Mul6&Mul7&Mul8&Mul9&Mul10\\
\hline

\multicolumn{2}{|c|}{ $  \bf \big<{\frac{dN_{ch}}{d\eta} } \big>$} &21.3$\pm$0.6&16.5$\pm$0.5&13.5$\pm$0.4&11.5$\pm$0.3&10.1$\pm$0.3&8.45$\pm$0.25&6.72$\pm$0.21&5.40$\pm$0.17&3.90$\pm$0.14&2.26$\pm$0.12\\
\hline
\end{tabular}
\end{adjustbox}
 \end{table*}
 


\begin{table*}[htbp]
\caption[p]{Number of mean charged particle multiplicity density for $\pi, K, p $ at mid rapidity corresponding to different event classes at $\sqrt{s}$ = 13 TeV~\cite{ALICE:2020nkc}.The uncertainties are the quadratic sum of statistical and systematic contributions.} 
\label{table:mult_info}
\begin{adjustbox}{max width=\textwidth}
\begin{tabular}{c|c|c|c|c|c|c|c|c|c|c|c|}
\hline
\multicolumn{2}{|c|}{${\bf Multiplicity class}$}&Mul1&Mul2&Mul3&Mul4&Mul5&Mul6&Mul7&Mul8&Mul9&Mul10\\
\hline

\multicolumn{2}{|c|}{ $  \bf \big<{\frac{dN_{ch}}{d\eta} } \big>$} &26.02$\pm$0.35&20.02$\pm$0.27&16.17$\pm$0.22&13.77$\pm$0.19&12.04$\pm$0.17&10.02$\pm$0.14&7.95$\pm$0.11&6.32$\pm$0.09&4.50$\pm$0.07&2.55 $\pm$0.04\\
\hline
\end{tabular}
\end{adjustbox}
 \end{table*}
 

 
\begin{table*}[htbp]
\caption[p]{Number of mean charged particle multiplicity density for $\phi,\Lambda, \Xi, \Omega $ at mid rapidity corresponding to different event classes at $\sqrt{s}$ = 13 TeV~\cite{ALICE:2019etb,ALICE:2019avo}. The uncertainties are the quadratic sum of statistical and systematic contributions.} 
\label{table:mult_info}
\begin{adjustbox}{max width=\textwidth}
\begin{tabular}{c|c|c|c|c|c|c|c|c|c|c|c|}
\hline
\multicolumn{2}{|c|}{${\bf Multiplicity class}$}&Mul1&Mul2&Mul3&Mul4&Mul5&Mul6&Mul7&Mul8&Mul9&Mul10\\
\hline

\multicolumn{2}{|c|}{ $  \bf \big<{\frac{dN_{ch}}{d\eta} } \big>$} &25.75$\pm$0.40&19.83$\pm$0.30&16.12$\pm$0.24&13.76$\pm$0.21&12.06$\pm$0.18&10.11$\pm$0.15&8.07$\pm$0.12&6.48$\pm$0.10&4.64$\pm$0.07&2.52 $\pm$0.04\\
\hline
\end{tabular}
\end{adjustbox}
 \end{table*}
 



\begin{table*}[htbp]
\caption[p]{{Fit results} at $\sqrt{s}$ = 7 TeV~\cite{ALICE:2018pal}, using data from the {ALICE}
Collaboration using Eqns.~\ref{YieldIndexZero} and~\ref{R0}.The reason of merging bins  Mul[1+2], Mul[3+4] etc for $\Omega$ is lack of statistics.}
\label{table:parameters}
\begin{adjustbox}{max width=\textwidth}
\begin{tabular}{|c|c|c|c|c|c|c|c|c|c|c|c|}

\hline
\multicolumn{2}{|c|}{${\bf Particles}$}&\multicolumn{10}{c|}{\bf Multiplicity class} \\
\cline{3-12}
\multicolumn{2}{|c|}{} &Mul1&Mul2&Mul3&Mul4&Mul5&Mul6&Mul7&Mul8&Mul9&Mul10\\


\hline
\multirow{4}{*}{$\pi^+ +\pi^-$} & $T_0$ (GeV)  &0.086$\pm$0.001&0.083$\pm$0.001&0.081$\pm$0.001&0.080$\pm$0.001&0.079$\pm$0.001&0.077$\pm$0.001&0.076$\pm$0.001&0.073$\pm$0.001&0.073$\pm$0.001&0.070$\pm$0.001\\
\cline{2-12}
                   
                    & $R_0$ (fm) &6.878$\pm$0.008 & 6.583$\pm$0.083 & 6.354$\pm$0.006&6.188$\pm$0.094 &6.042$\pm$0.066&5.855$\pm$0.007&5.596$\pm$0.077&5.454$\pm$0.010&5.021$\pm$0.083&4.553$\pm$0.090\\

\cline{2-12}
                   
                    & $q$ &1.170$\pm$0.001 & 1.168$\pm$0.001 & 1.167$\pm$0.001&1.165$\pm$0.001 &1.164$\pm$0.001&1.162$\pm$0.001&1.160$\pm$0.001&1.158$\pm$0.001&1.151$\pm$0.001&1.139$\pm$0.001\\
                    \cline{2-12}
                       & $\chi^2$/ndf & 6.923&7.184&6.383&5.587&4.981&4.117&2.980&1.967&0.583&0.117\\                      
                      
          
\hline
\multirow{4}{*}{$K^+ +K^-$}& $T_0$ (GeV) &0.154$\pm$0.004&0.140$\pm$0.004&0.130$\pm$0.003&0.122$\pm$0.003&0.116$\pm$0.003&0.109$\pm$0.003&0.100$\pm$0.003&0.092$\pm$0.003&0.080$\pm$0.003&0.061$\pm$0.003\\

\cline{2-12}
 		& $R_0$ (fm) &2.394$\pm$0.072 & 2.467$\pm$0.076 & 2.523$\pm$0.082&2.577$\pm$0.088&2.645$\pm$0.096&2.687$\pm$0.101&2.795$\pm$0.113&2.933$\pm$0.129&3.181$\pm$0.168&4.144$\pm$0.323\\
                    
\cline{2-12}                  
                    & $q$ &1.137$\pm$0.003 & 1.141$\pm$0.002 & 1.143$\pm$0.002&1.145$\pm$0.002 &1.147$\pm$0.002&1.148$\pm$0.002&1.150$\pm$0.002&1.151$\pm$0.002&1.152$\pm$0.002&1.149$\pm$0.002\\
                    \cline{2-12}
                    & $\chi^2$/ndf & 0.087&0.077&0.081&0.085&0.081&0.078&0.053&0.048&0.038&0.114\\
                    
\hline
\multirow{4}{*}{$p +\bar{p}$} & $T_0$ (fm)  &0.205$\pm$0.008&0.183$\pm$0.008&0.163$\pm$0.007&0.146$\pm$0.007&0.137$\pm$0.007&0.121$\pm$0.006&0.104$\pm$0.006&0.086$\pm$0.006&0.064$\pm$0.006&0.026$\pm$0.006\\
\cline{2-12}
                   
                    & $R_0$ (fm) &1.339$\pm$0.070 & 1.472$\pm$0.082 & 1.643$\pm$0.102&1.845$\pm$0.128 &1.973$\pm$0.145&2.271$\pm$0.186&2.737$\pm$0.275&3.600$\pm$0.003&5.699$\pm$0.009&32.505$\pm$0.010\\

\cline{2-12}
                   
                    & $q$ &1.097$\pm$0.004 & 1.100$\pm$0.004 & 1.104$\pm$0.004&1.108$\pm$0.004 &1.109$\pm$0.004&1.112$\pm$0.004&1.116$\pm$0.004&1.120$\pm$0.004&1.124$\pm$0.004&1.129$\pm$0.004\\
                    \cline{2-12}
                 & $\chi^2$/ndf& 0.079&0.079&0.078&0.085&0.085&0.100&0.092&0.090&0.074&0.114\\

\hline
\multirow{3}{*}{$\Lambda+\bar{\Lambda}$}&$T_0$ (GeV) &0.253$\pm$ 0.005&0.217$\pm$ 0.004&0.186$\pm$0.004&0.165$\pm$0.003&0.151$\pm$0.008&0.137$\pm$0.001&0.107$\pm$0.003&0.079$\pm$0.002&0.054$\pm$0.006&0.014$\pm$0.001\\

\cline{2-12}
                   
                    & $R_0$ (fm) &0.790$\pm$0.008 & 0.906$\pm$0.005 & 1.058$\pm$0.006&1.201$\pm$0.003 &1.334$\pm$0.005&1.480$\pm$0.007&2.068$\pm$0.004&3.330$\pm$0.008&6.245$\pm$0.007&100.545$\pm$7.691\\
                   
                   \cline{2-12} 
                    & $q$  & 1.081$\pm$0.004& 1.089$\pm$0.005& 1.097$\pm$0.002&1.102$\pm$0.004&1.104$\pm$0.004&1.106$\pm$0.001&1.115$\pm$0.006&1.124$\pm$0.001&1.128$\pm$0.005&1.136$\pm$0.001\\
                    \cline{2-12}
                    & $\chi^2$/ndf& 0.419&0.184&0.161&0.176&0.197&0.121&0.053&0.129&0.120&0.091\\
                    
\hline
\multirow{3}{*}{$\Xi^{-}+\bar{\Xi}^{+}$}& $T_0$ (GeV) &0.315$\pm$0.004&0.267$\pm$0.010&0.224$\pm$0.007&0.211$\pm$0.005&0.185$\pm$0.003&0.164$\pm$0.007&0.146$\pm$0.003&0.103$\pm$0.001&0.073$\pm$0.002 &0.021$\pm$0.007\\

\cline{2-12}
                   
                    & $R_0$ (fm) &0.405$\pm$0.004 & 0.463$\pm$0.010 & 0.569$\pm$0.007&0.582$\pm$0.083 &0.675$\pm$0.008&0.775$\pm$0.007&0.868$\pm$0.024&1.470$\pm$0.013&2.511$\pm$0.392&32.494$\pm$1.897\\
                   
                   \cline{2-12} 
                    & $q$  &1.067$\pm$0.006 & 1.078$\pm$0.002&1.086$\pm$0.003 &1.088$\pm$0.002 &1.096$\pm$0.005&1.100$\pm$0.007&1.102$\pm$0.005&1.115$\pm$0.001&1.121$\pm$0.004&1.132$\pm$0.002\\
                  \cline{2-12}
                    & $\chi^2$/ndf& 0.418&0.402&0.298&0.114&0.149&0.196&0.201&0.571&0.267&0.168\\
                    
 \hline 
\multicolumn{2}{c|}{} & Mul1 & Mul2 & Mul3 &\multicolumn{2}{c|}{ Mul[4+5]} & Mul6 &  Mul7 &  Mul8 & Mul9& Mul10\\
\hline
\multirow{3}{*}{$\phi$} & $T_0$ (GeV)  &0.270$\pm$0.002&0.236$\pm$0.004&0.216$\pm$0.006&\multicolumn{2}{c|}{0.192$\pm$0.001}&0.159$\pm$0.001&0.135$\pm$0.001&0.112$\pm$0.006&0.082$\pm$0.001&0.025$\pm$0.008\\
\cline{2-12}
                   
                    & $R_0$ (fm) &0.540$\pm$0.009 & 0.591$\pm$0.007 & 0.630$\pm$0.007&\multicolumn{2}{c|}{0.680$\pm$0.005}&0.817$\pm$0.006&0.955$\pm$0.007&1.173$\pm$0.004&1.740$\pm$0.010&15.414$\pm$0.069\\

\cline{2-12}
                   
                    & $q$ &1.103$\pm$0.004 & 1.109$\pm$0.005 & 1.112$\pm$0.002&\multicolumn{2}{c|}{0.119$\pm$0.003}&1.126$\pm$0.002&1.134$\pm$0.004&1.140$\pm$0.002&1.146$\pm$0.002&1.155$\pm$0.002\\
                    \cline{2-12}
                    & $\chi^2$/ndf& 1.046&0.318&0.429&\multicolumn{2}{c|}{0.280}&0.585&0.335&0.525&0.543&0.220\\
                      
\hline 
\multicolumn{2}{c|}{}&\multicolumn{2}{c|}{ Mul[1+2]}&\multicolumn{2}{c|}{ Mul[3+4]}&\multicolumn{2}{c|}{ Mul[5+6]}&\multicolumn{2}{c|}{ Mul[7+8]}&\multicolumn{2}{c|}{ Mul[9+10]}\\
\hline
\multirow{3}{*}{$\Omega^{-}+\bar{\Omega}^{+}$}& $T_0$ (GeV)  &\multicolumn{2}{c|}{0.360$\pm$0.006}&\multicolumn{2}{c|}{0.309$\pm$0.007}&\multicolumn{2}{c|}{0.242$\pm$0.010}&\multicolumn{2}{c|}{0.122$\pm$0.007}&\multicolumn{2}{c|}{0.036$\pm$0.001}\\

\cline{2-12} 
  &  $R_0$ (fm)  &\multicolumn{2}{c|}{0.137$\pm$0.006} & \multicolumn{2}{c|}{0.145$\pm$0.006}&\multicolumn{2}{c|}{0.188$\pm$0.005}&\multicolumn{2}{c|}{0.503$\pm$0.009}&\multicolumn{2}{c|}{4.696$\pm$0.009} \\

 \cline{2-12} 
  & $q$ &\multicolumn{2}{c|}{1.056$\pm$0.009} & \multicolumn{2}{c|}{1.065$\pm$0.001}&\multicolumn{2}{c|}{1.076$\pm$0.005}&\multicolumn{2}{c|}{1.110$\pm$0.009}&\multicolumn{2}{c|}{1.130$\pm$0.007} \\
 \cline{2-12}
 & $\chi^2$/ndf& \multicolumn{2}{c|}{0.362}&\multicolumn{2}{c|}{0.450}&\multicolumn{2}{c|}{0.411}&\multicolumn{2}{c|}{0.236}&\multicolumn{2}{c|}{0.580}\\
 \hline 
 \end{tabular}

\end{adjustbox}
\label{ALICE_T0_7}
 \end{table*}


 
\begin{table*}[htbp]
\caption[p]{{Fit results} at $\sqrt{s}$ = 13 TeV~\cite{ALICE:2020nkc,ALICE:2019etb,ALICE:2019avo}, using data from the {ALICE}
Collaboration using Eqns.~\ref{YieldIndexZero} and~\ref{R0}. The reason of merging bins  Mul[1+2], Mul[3+4] etc for $\Omega$ is lack of statistics.}
\label{table:parameters}
\begin{adjustbox}{max width=\textwidth}
\begin{tabular}{|c|c|c|c|c|c|c|c|c|c|c|c|}

\hline
\multicolumn{2}{|c|}{${\bf Particles}$}&\multicolumn{10}{c|}{\bf Multiplicity class} \\
\cline{3-12}
\multicolumn{2}{|c|}{} &Mul1&Mul2&Mul3&Mul4&Mul5&Mul6&Mul7&Mul8&Mul9&Mul10\\
\hline
\multirow{4}{*}{$\pi^+ +\pi^-$} & $T_0$ (GeV)  &0.089$\pm$0.001&0.087$\pm$0.001&0.084$\pm$0.001&0.083$\pm$0.001&0.081$\pm$0.001&0.079$\pm$0.001&0.078$\pm$0.001&0.076$\pm$0.001&0.074$\pm$0.001&0.070$\pm$0.001\\
\cline{2-12}
                   
                    & $R_0$ (fm) &6.989$\pm$0.006 & 6.652$\pm$0.006 & 6.422$\pm$0.005&6.250$\pm$0.003 &6.105$\pm$0.006&5.922$\pm$0.006&5.671$\pm$0.006&5.431$\pm$0.006&5.068$\pm$0.006&4.585$\pm$0.008\\

\cline{2-12}
                   
                    & $q$ &1.176$\pm$0.001 & 1.173$\pm$0.001 & 1.171$\pm$0.001&1.170$\pm$0.001 &1.169$\pm$0.001&1.167$\pm$0.001&1.165$\pm$0.001&1.162$\pm$0.001&1.157$\pm$0.001&1.145$\pm$0.0\\
                    \cline{2-12}
                       & $\chi^2$/ndf & 7.959&7.130&6.254&5.563&5.006&4.158&3.199&2.217&1.034&0.504\\                      
                      
\hline
\multirow{4}{*}{$K^+ +K^-$}& $T_0$ (GeV) &0.161$\pm$0.001&0.145$\pm$0.003&0.134$\pm$0.003&0.126$\pm$0.003&0.120$\pm$0.003&0.113$\pm$0.003&0.103$\pm$0.003&0.095$\pm$0.003&0.082$\pm$0.002&0.055$\pm$0.0\\

\cline{2-12}
 		& $R_0$ (fm) &2.328$\pm$0.007 & 2.462$\pm$0.070 & 2.515$\pm$0.075&2.569$\pm$0.080&2.609$\pm$0.085&2.666$\pm$0.092&2.775$\pm$0.104&2.858$\pm$0.114&3.126$\pm$0.139&4.815$\pm$0.014\\
                    
\cline{2-12}                  
                    & $q$ &1.145$\pm$0.001 & 1.149$\pm$0.002 & 1.151$\pm$0.002&1.152$\pm$0.002 &1.154$\pm$0.002&1.155$\pm$0.002&1.156$\pm$0.002&1.157$\pm$0.002&1.158$\pm$0.002&1.160$\pm$0.001\\
                    \cline{2-12}
                    & $\chi^2$/ndf & 0.585&0.154&0.143&0.153&0.201&0.199&0.156&0.204&0.136&0.559\\
\hline
\multirow{4}{*}{$p +\bar{p}$} & $T_0$ (fm)  &0.228$\pm$0.007&0.197$\pm$0.007&0.172$\pm$0.007&0.153$\pm$0.007&0.140$\pm$0.007&0.123$\pm$0.007&0.104$\pm$0.006&0.084$\pm$0.007&0.059$\pm$0.001&0.017$\pm$0.001\\
\cline{2-12}
                   
                    & $R_0$ (fm) &1.227$\pm$0.006 & 1.348$\pm$0.013 & 1.563$\pm$0.098&1.768$\pm$0.124 &1.935$\pm$0.148&2.250$\pm$0.198&2.785$\pm$0.0075&3.700$\pm$0.488&6.499$\pm$1.333&79.389$\pm$0.006\\

\cline{2-12}
                   
                    & $q$ &1.098$\pm$0.004 & 1.104$\pm$0.004 & 1.109$\pm$0.004&1.114$\pm$0.004 &1.116$\pm$0.004&1.120$\pm$0.004&1.123$\pm$0.004&1.128$\pm$0.004&1.133$\pm$0.004&1.139$\pm$0.004\\
                    \cline{2-12}
                 & $\chi^2$/ndf& 0.170&0.151&0.168&0.152&0.137&0.158&0.142&0.119&0.166&0.192\\ 
                 

\hline
\multirow{4}{*}{$\phi$} & $T_0$ (fm)  &0.310$\pm$0.003&0.266$\pm$0.005&0.251$\pm$0.002&0.224$\pm$0.005&0.205$\pm$0.013&0.182$\pm$0.004&0.155$\pm$0.006&0.140$\pm$0.006&0.108$\pm$0.001&0.049$\pm$0.001\\
\cline{2-12}
                   
                    & $R_0$ (fm) &0.481$\pm$0.006 & 0.533$\pm$0.003 & 0.535$\pm$0.006&0.570$\pm$0.010 &0.618$\pm$0.006&0.691$\pm$0.005&0.796$\pm$0.007&0.832$\pm$0.004&1.108$\pm$0.007&3.505$\pm$0.009\\

\cline{2-12}
                   
                    & $q$ &1.097$\pm$0.004 & 1.105$\pm$0.006 & 1.108$\pm$0.003&1.119$\pm$0.006 &1.124$\pm$0.002&1.126$\pm$0.005&1.135$\pm$0.005&1.140$\pm$0.005&1.146$\pm$0.002&1.154$\pm$0.006\\
                    \cline{2-12}
                 & $\chi^2$/ndf& 0.721&0.491&0.396&0.584&0.432&0.371&0.497&0.215&0.510&0.763\\

\hline
\multirow{3}{*}{$\Lambda+\bar{\Lambda}$}&$T_0$ (GeV) &0.286$\pm$ 0.005&0.246$\pm$ 0.006&0.218$\pm$0.006&0.187$\pm$0.004&0.176$\pm$0.006&0.152$\pm$0.003&0.121$\pm$0.002&0.101$\pm$0.006&0.069$\pm$0.004&0.018$\pm$0.005\\

\cline{2-12}
                   
                    & $R_0$ (fm) &0.705$\pm$0.005 & 0.800$\pm$0.007 & 0.889$\pm$0.005&1.050$\pm$0.003 &1.098$\pm$0.008&1.308$\pm$0.150&1.740$\pm$0.005&2.204$\pm$0.007&4.019$\pm$0.008&6.184$\pm$0.255\\
                   
                   \cline{2-12} 
                    & $q$  & 1.081$\pm$0.007& 1.088$\pm$0.007& 1.093$\pm$0.004&1.102$\pm$0.007&1.103$\pm$0.007&1.109$\pm$0.007&1.117$\pm$0.001&1.122$\pm$0.007&1.128$\pm$0.002&1.138$\pm$0.001\\
                    \cline{2-12}
                    & $\chi^2$/ndf& 0.560&0.355&0.373&0.433&0.338&0.230&0.333&0.178&0.229&0.260\\
                    
\hline
\multirow{3}{*}{$\Xi^{-}+\bar{\Xi}^{+}$}& $T_0$ (GeV) &0.347$\pm$0.007&0.309$\pm$0.002&0.253$\pm$0.007&0.237$\pm$0.006&0.199$\pm$0.006&0.176$\pm$0.001&0.158$\pm$0.005&0.127$\pm$0.001&0.102$\pm$0.005 &0.026$\pm$0.001\\

\cline{2-12}
                   
                    & $R_0$ (fm) &0.383$\pm$0.008 & 0.405$\pm$0.005 & 0.495$\pm$0.006&0.515$\pm$0.007 &0.630$\pm$0.008&0.722$\pm$0.008&0.796$\pm$0.003&1.079$\pm$0.005&1.330$\pm$0.007&1.784$\pm$0.040\\
                   
                   \cline{2-12} 
                    & $q$  &1.068$\pm$0.009 & 1.074$\pm$0.005&1.089$\pm$0.008 &1.091$\pm$0.007 &1.103$\pm$0.007&1.106$\pm$0.002&1.108$\pm$0.005&1.113$\pm$0.002&1.120$\pm$0.006&1.137$\pm$0.002\\
                  \cline{2-12}
                    & $\chi^2$/ndf& 0.430&0.229&0.434&0.450&0.527&0.232&0.380&0.535&0.305&0.305\\
                      
\hline 
\multicolumn{2}{c|}{}&\multicolumn{2}{c|}{ Mul[1+2]}&\multicolumn{2}{c|}{ Mul[3+4]}&\multicolumn{2}{c|}{ Mul[5+6]}&\multicolumn{2}{c|}{ Mul[7+8]}&\multicolumn{2}{c|}{ Mul[9+10]}\\
\hline
\multirow{3}{*}{$\Omega^{-}+\bar{\Omega}^{+}$}& $T_0$ (GeV)  &\multicolumn{2}{c|}{0.402$\pm$0.007}&\multicolumn{2}{c|}{0.311$\pm$0.002}&\multicolumn{2}{c|}{0.265$\pm$0.005}&\multicolumn{2}{c|}{0.198$\pm$0.007}&\multicolumn{2}{c|}{0.130$\pm$0.009}\\

\cline{2-12} 
  &  $R_0$ (fm)  &\multicolumn{2}{c|}{0.626$\pm$0.030} & \multicolumn{2}{c|}{0.826$\pm$0.005}&\multicolumn{2}{c|}{0.847$\pm$0.006}&\multicolumn{2}{c|}{1.107$\pm$0.005}&\multicolumn{2}{c|}{1.671$\pm$0.006} \\

 \cline{2-12} 
  & $q$ &\multicolumn{2}{c|}{1.057$\pm$0.009} & \multicolumn{2}{c|}{1.065$\pm$0.010}&\multicolumn{2}{c|}{1.085$\pm$0.011}&\multicolumn{2}{c|}{1.097$\pm$0.006}&\multicolumn{2}{c|}{1.106$\pm$0.005} \\
 \cline{2-12}
 & $\chi^2$/ndf& \multicolumn{2}{c|}{0.048}&\multicolumn{2}{c|}{1.625}&\multicolumn{2}{c|}{0.430}&\multicolumn{2}{c|}{1.525}&\multicolumn{2}{c|}{0.729}\\
 \hline 
 \end{tabular}

\end{adjustbox}
\label{ALICE_T0_13}
 \end{table*}



\begin{table*}[htbp]
\caption[p]{{Fit results} at $\sqrt{s}$ = 7 TeV~\cite{ALICE:2018pal}, using data from the
ALICE Collaboration with $q$ from Table~\ref{ALICE_T0_7} following Eqns.~\ref{YieldNonZeroMu} and~\ref{R}.} 
\label{table:parameters}
\begin{adjustbox}{max width=\textwidth}
\begin{tabular}{|c|c|c|c|c|c|c|c|c|c|c|c|}

\hline
\multicolumn{2}{|c|}{${\bf Particles}$}&\multicolumn{10}{c|}{\bf Multiplicity class} \\
\cline{3-12}
\multicolumn{2}{|c|}{} &Mul1&Mul2&Mul3&Mul4&Mul5&Mul6&Mul7&Mul8&Mul9&Mul10\\
\hline
\multirow{4}{*}{$\pi^+ +\pi^-$} & $T$ (GeV)  &0.196$\pm$0.006&0.193$\pm$0.003&0.189$\pm$0.007&0.186$\pm$0.007&0.182$\pm$0.008&0.175$\pm$0.002&0.169$\pm$0.008&0.163$\pm$0.005&0.153$\pm$0.001&0.134$\pm$0.007\\
\cline{2-12}
                   
                    & $R$ (fm) &1.060$\pm$0.008 & 0.959$\pm$0.007 & 0.900$\pm$0.007&0.861$\pm$0.006 &0.851$\pm$0.005&0.848$\pm$0.008&0.811$\pm$0.004&0.788$\pm$0.006&0.787$\pm$0.007&0.785$\pm$0.005\\

\cline{2-12}
                   
                    & $\mu$ &0.642$\pm$0.002 & 0.649$\pm$0.007 & 0.644$\pm$0.005&0.640$\pm$0.007 &0.624$\pm$0.006&0.599$\pm$0.005&0.584$\pm$0.008&0.564$\pm$0.005&0.523$\pm$0.009&0.459$\pm$0.005\\
                    \cline{2-12}
                       & $\chi^2$/ndf & 6.928&7.194&6.383&5.597&4.984&4.130&2.980&1.953&0.595&0.117\\                      
                      
\hline
\multirow{4}{*}{$K^+ +K^-$}& $T$ (GeV) &0.197$\pm$0.009&0.193$\pm$0.005&0.189$\pm$0.006&0.186$\pm$0.006&0.182$\pm$0.009&0.175$\pm$0.005&0.169$\pm$0.009&0.164$\pm$0.009&0.149$\pm$0.007&0.124$\pm$0.009\\

\cline{2-12}
 		& $R$ (fm) &1.227$\pm$0.005 & 1.046$\pm$0.005 & 0.941$\pm$0.005&0.871$\pm$0.007&0.821$\pm$0.005&0.802$\pm$0.008&0.736$\pm$0.007&0.675$\pm$0.006&0.672$\pm$0.006&0.671$\pm$0.006\\
                    
\cline{2-12}                  
                    & $\mu$ &0.308$\pm$0.005 & 0.372$\pm$0.007 & 0.407$\pm$0.007&0.430$\pm$0.006 &0.447$\pm$0.005&0.441$\pm$0.007&0.459$\pm$0.007&0.476$\pm$0.006&0.451$\pm$0.007&0.423$\pm$0.005\\
                    \cline{2-12}
                    & $\chi^2$/ndf & 0.087&0.077&0.082&0.087&0.083&0.080&0.053&0.050&0.038&0.115\\
\hline
\multirow{4}{*}{$p +\bar{p}$} & $T$ (fm)  &0.198$\pm$0.001&0.193$\pm$0.006&0.189$\pm$0.006&0.187$\pm$0.005&0.183$\pm$0.006&0.176$\pm$0.005&0.169$\pm$0.006&0.163$\pm$0.007&0.149$\pm$0.007&0.117$\pm$0.006\\
\cline{2-12}
                   
                    & $R$ (fm) &1.546$\pm$0.005 & 1.214$\pm$0.005 & 0.984$\pm$0.006&0.794$\pm$0.008 &0.739$\pm$0.005&0.676$\pm$0.007&0.578$\pm$0.005&0.491$\pm$0.006&0.448$\pm$0.008&0.446$\pm$0.007\\

\cline{2-12}
                   
                    & $\mu$ &-0.082$\pm$0.008 & 0.096$\pm$0.005 & 0.242$\pm$0.006&0.377$\pm$0.005 &0.421$\pm$0.006&0.477$\pm$0.007&0.562$\pm$0.005&0.642$\pm$0.008&0.681$\pm$0.005&0.696$\pm$0.005\\
                    \cline{2-12}
                 & $\chi^2$/ndf& 0.079&0.080&0.079&0.085&0.085&0.101&0.092&0.090&0.074&0.114\\

\hline
\multirow{3}{*}{$\Lambda+\bar{\Lambda}$}&$T$ (GeV) &0.201$\pm$ 0.007&0.196$\pm$ 0.008&0.191$\pm$0.006&0.187$\pm$0.006&0.184$\pm$0.005&0.180$\pm$0.007&0.171$\pm$0.007&0.163$\pm$0.007&0.153$\pm$0.009&0.119$\pm$0.009\\

\cline{2-12}
                   
                    & $R$ (fm) &2.226$\pm$0.005 & 1.380$\pm$0.008 & 0.964$\pm$0.007&0.772$\pm$0.008 &0.675$\pm$0.007&0.586$\pm$0.006&0.466$\pm$0.005&0.378$\pm$0.006&0.309$\pm$0.007&0.299$\pm$0.007\\
                   
                   \cline{2-12} 
                    & $\mu$  & -0.653$\pm$0.006& -0.239$\pm$0.007& 0.046$\pm$0.005&0.211$\pm$0.007&0.303$\pm$0.006&0.393$\pm$0.007&0.584$\pm$0.005&0.671$\pm$0.007&0.763$\pm$0.009&0.770$\pm$0.005\\
                    \cline{2-12}
                    & $\chi^2$/ndf& 0.419&0.184&0.162&0.176&0.199&0.124&0.053&0.131&0.116&0.091\\
                    
\hline
\multirow{3}{*}{$\Xi^{-}+\bar{\Xi}^{+}$}& $T$ (GeV) &0.202$\pm$0.005&0.197$\pm$0.002&0.194$\pm$0.007&0.188$\pm$0.007&0.185$\pm$0.006&0.179$\pm$0.007&0.171$\pm$0.007&0.164$\pm$0.007&0.156$\pm$0.008 &0.122$\pm$0.007\\

\cline{2-12}
                   
                    & $R$ (fm) &4.307$\pm$0.005 & 1.907$\pm$0.005 & 1.040$\pm$0.008&0.954$\pm$0.007 &0.678$\pm$0.007&0.564$\pm$0.005&0.488$\pm$0.007&0.332$\pm$0.006&0.249$\pm$0.006&0.222$\pm$0.007\\
                   
                   \cline{2-12} 
                    & $\mu$  &-1.698$\pm$0.006 & -0.916$\pm$0.005&-0.353$\pm$0.006 &-0.276$\pm$0.006 &-0.005$\pm$0.005&0.146$\pm$0.005&0.249$\pm$0.007&0.521$\pm$0.005&0.670$\pm$0.007&0.757$\pm$0.009\\
                  \cline{2-12}
                    & $\chi^2$/ndf& 0.418&0.409&0.299&0.115&0.149&0.196&0.201&0.562&0.268&0.168\\
                    
 \hline 
\multicolumn{2}{c|}{} & Mul1 & Mul2 & Mul3 &\multicolumn{2}{c|}{ Mul[4+5]} & Mul6 &  Mul7 &  Mul8 & Mul9& Mul10\\
\hline
\multirow{3}{*}{$\phi$} & $T$ (GeV)  &0.198$\pm$0.007&0.193$\pm$0.007&0.189$\pm$0.005&\multicolumn{2}{c|}{0.183$\pm$0.005}&0.175$\pm$0.005&0.171$\pm$0.006&0.163$\pm$0.005&0.149$\pm$0.005&0.118$\pm$0.001\\
\cline{2-12}
                   
                    & $R$ (fm) &1.610$\pm$0.005 & 1.186$\pm$0.005 & 0.980$\pm$0.007&\multicolumn{2}{c|}{0.797$\pm$0.006}&0.631$\pm$0.005&0.491$\pm$0.005&0.421$\pm$0.005&0.375$\pm$0.006&0.333$\pm$0.008\\

\cline{2-12}
                   
                    & $\mu$ &-0.687$\pm$0.005 & -0.410$\pm$0.005 & -0.241$\pm$0.007&\multicolumn{2}{c|}{-0.080$\pm$0.010}&0.101$\pm$0.007&0.266$\pm$0.007&0.365$\pm$0.006&0.450$\pm$0.006&0.596$\pm$0.009\\
                    \cline{2-12}
                    & $\chi^2$/ndf& 1.046&0.318&0.429&\multicolumn{2}{c|}{0.281}&0.550&0.342&0.527&0.549&0.233\\
                      
\hline 
\multicolumn{2}{c|}{}&\multicolumn{2}{c|}{ Mul[1+2]}&\multicolumn{2}{c|}{ Mul[3+4]}&\multicolumn{2}{c|}{ Mul[5+6]}&\multicolumn{2}{c|}{ Mul[7+8]}&\multicolumn{2}{c|}{ Mul[9+10]}\\
\hline
\multirow{3}{*}{$\Omega^{-}+\bar{\Omega}^{+}$}& $T$ (GeV)  &\multicolumn{2}{c|}{0.204$\pm$0.008}&\multicolumn{2}{c|}{0.199$\pm$0.007}&\multicolumn{2}{c|}{0.189$\pm$0.005}&\multicolumn{2}{c|}{0.165$\pm$0.006}&\multicolumn{2}{c|}{0.134$\pm$0.005}\\

\cline{2-12} 
  &  $R$ (fm)  &\multicolumn{2}{c|}{5.023$\pm$0.005} & \multicolumn{2}{c|}{1.650$\pm$0.008}&\multicolumn{2}{c|}{0.608$\pm$0.005}&\multicolumn{2}{c|}{0.186$\pm$0.007}&\multicolumn{2}{c|}{0.107$\pm$0.006} \\

 \cline{2-12} 
  & $\mu$ &\multicolumn{2}{c|}{-2.838$\pm$0.006} & \multicolumn{2}{c|}{-1.729$\pm$0.009}&\multicolumn{2}{c|}{-0.721$\pm$0.008}&\multicolumn{2}{c|}{0.382$\pm$0.005}&\multicolumn{2}{c|}{0.752$\pm$0.005} \\
 \cline{2-12}
 & $\chi^2$/ndf& \multicolumn{2}{c|}{0.354}&\multicolumn{2}{c|}{0.450}&\multicolumn{2}{c|}{0.400}&\multicolumn{2}{c|}{0.237}&\multicolumn{2}{c|}{0.580}\\
 \hline 
 \end{tabular}

\end{adjustbox} 
 \end{table*}


 
\begin{table*}[htbp]
\caption[p]{{Fit results} at $\sqrt{s}$ = 13 TeV~\cite{ALICE:2020nkc,ALICE:2019etb,ALICE:2019avo}, using data from the
ALICE Collaboration with $q$ from Table~\ref{ALICE_T0_13} following Eqns.~\ref{YieldNonZeroMu} and~\ref{R}.}
\label{table:parameters}
\begin{adjustbox}{max width=\textwidth}
\begin{tabular}{|c|c|c|c|c|c|c|c|c|c|c|c|}

\hline
\multicolumn{2}{|c|}{${\bf Particles}$}&\multicolumn{10}{c|}{\bf Multiplicity class} \\
\cline{3-12}
\multicolumn{2}{|c|}{} &Mul1&Mul2&Mul3&Mul4&Mul5&Mul6&Mul7&Mul8&Mul9&Mul10\\
\hline
\multirow{4}{*}{$\pi^+ +\pi^-$} & $T$ (GeV)  &0.200$\pm$0.007&0.194$\pm$0.003&0.190$\pm$0.007&0.186$\pm$0.002&0.182$\pm$0.007&0.177$\pm$0.007&0.170$\pm$0.006&0.162$\pm$0.007&0.150$\pm$0.003&0.131$\pm$0.007\\
\cline{2-12}
                   
                    & $R$ (fm) &1.175$\pm$0.005 & 1.093$\pm$0.006 & 1.028$\pm$0.006&0.984$\pm$0.008 &0.961$\pm$0.006&0.932$\pm$0.007&0.907$\pm$0.006&0.897$\pm$0.006&0.892$\pm$0.007&0.887$\pm$0.008\\

\cline{2-12}
                   
                    & $\mu$ &0.626$\pm$0.005 & 0.617$\pm$0.005 & 0.612$\pm$0.007&0.606$\pm$0.005 &0.595$\pm$0.005&0.581$\pm$0.005&0.557$\pm$0.007&0.529$\pm$0.007&0.486$\pm$0.006&0.419$\pm$0.007\\
                    \cline{2-12}
                       & $\chi^2$/ndf & 7.960&7.134&6.264&5.566&5.007&4.165&3.200&2.220&1.035&0.505\\                      
                      
\hline
\multirow{4}{*}{$K^+ +K^-$}& $T$ (GeV) &0.200$\pm$0.008&0.194$\pm$0.005&0.190$\pm$0.006&0.186$\pm$0.006&0.182$\pm$0.007&0.177$\pm$0.009&0.170$\pm$0.007&0.162$\pm$0.006&0.151$\pm$0.007&0.131$\pm$0.005\\

\cline{2-12}
 		& $R$ (fm) &1.310$\pm$0.007 & 1.174$\pm$0.007 & 1.043$\pm$0.005&0.982$\pm$0.006&0.934$\pm$0.005&0.879$\pm$0.008&0.817$\pm$0.005&0.778$\pm$0.008&0.722$\pm$0.005&0.609$\pm$0.006\\
                    
\cline{2-12}                  
                    & $\mu$ &0.274$\pm$0.005&0.326$\pm$0.007&0.369$\pm$0.006&0.384$\pm$0.007&0.398$\pm$0.008&0.410$\pm$0.006&0.423$\pm$0.006&0.425$\pm$0.007&0.430$\pm$0.006&0.462$\pm$0.005\\
                    \cline{2-12}
                    & $\chi^2$/ndf & 0.585&0.154&0.143&0.157&0.201&0.201&0.160&0.204&0.138&0.424\\
\hline
\multirow{4}{*}{$p +\bar{p}$} & $T$ (fm)  &0.202$\pm$0.006&0.196$\pm$0.003&0.192$\pm$0.007&0.188$\pm$0.007&0.183$\pm$0.007&0.178$\pm$0.007&0.171$\pm$0.007&0.163$\pm$0.005&0.150$\pm$0.006&0.130$\pm$0.005\\
\cline{2-12}
                   
                    & $R$ (fm) &1.981$\pm$0.006 & 1.413$\pm$0.005 & 1.107$\pm$0.005&0.917$\pm$0.005 &0.840$\pm$0.008&0.718$\pm$0.006&0.620$\pm$0.006&0.534$\pm$0.006&0.457$\pm$0.006&0.331$\pm$0.005\\

\cline{2-12}
                   
                    & $\mu$ &-0.286$\pm$0.007 & -0.148$\pm$0.005 & 0.164$\pm$0.006&0.299$\pm$0.005 &0.358$\pm$0.006&0.455$\pm$0.005&0.537$\pm$0.008&0.616$\pm$0.006&0.685$\pm$0.005&0.804$\pm$0.005\\
                    \cline{2-12}
                 & $\chi^2$/ndf& 0.170&0.151&0.169&0.152&0.138&0.159&0.144&0.119&0.167&0.193\\ 
                 

\hline
\multirow{4}{*}{$\phi$} & $T$ (fm)  &0.203$\pm$0.003&0.196$\pm$0.007&0.193$\pm$0.007&0.188$\pm$0.007&0.183$\pm$0.005&0.178$\pm$0.006&0.173$\pm$0.006&0.164$\pm$0.008&0.152$\pm$0.007&0.126$\pm$0.009\\
\cline{2-12}
                   
                    & $R$ (fm) &2.385$\pm$0.006 & 1.578$\pm$0.005 & 1.314$\pm$0.008&0.978$\pm$0.006 &0.867$\pm$0.005&0.739$\pm$0.005&0.587$\pm$0.007&0.531$\pm$0.006&0.451$\pm$0.005&0.342$\pm$0.006\\

\cline{2-12}
                   
                    & $\mu$ &-1.113$\pm$0.006 & -0.679$\pm$0.008 & -0.540$\pm$0.006&-0.291$\pm$0.006 &-0.173$\pm$0.007&-0.032$\pm$0.006&0.129$\pm$0.006&0.183$\pm$0.005&0.307$\pm$0.005&0.505$\pm$0.006\\
                    \cline{2-12}
                 & $\chi^2$/ndf& 0.721&0.491&0.395&0.578&0.434&0.371&0.497&0.213&0.507&0.750\\

\hline
\multirow{3}{*}{$\Lambda+\bar{\Lambda}$}&$T$ (GeV) &0.205$\pm$ 0.007&0.198$\pm$ 0.007&0.194$\pm$0.004&0.190$\pm$0.007&0.185$\pm$0.005&0.180$\pm$0.005&0.175$\pm$0.006&0.166$\pm$0.005&0.154$\pm$0.006&0.135$\pm$0.009\\

\cline{2-12}
                   
                    & $R$ (fm) &3.070$\pm$0.005 & 1.976$\pm$0.006 & 1.403$\pm$0.007&0.998$\pm$0.006 &0.934$\pm$0.006&0.740$\pm$0.006&0.554$\pm$0.006&0.486$\pm$0.006&0.394$\pm$0.007&0.274$\pm$0.006\\
                   
                   \cline{2-12} 
                    & $\mu$  & -0.998$\pm$0.006& -0.557$\pm$0.006& -0.261$\pm$0.007&0.022$\pm$0.005&0.072$\pm$0.005&0.252$\pm$0.007&0.445$\pm$0.005&0.530$\pm$0.006&0.649$\pm$0.005&0.844$\pm$0.006\\
                    \cline{2-12}
                    & $\chi^2$/ndf& 0.560&0.356&0.373&0.434&0.340&0.231&0.335&0.178&0.231&0.260\\
                    
\hline
\multirow{3}{*}{$\Xi^{-}+\bar{\Xi}^{+}$}& $T$ (GeV) &0.206$\pm$0.002&0.200$\pm$0.007&0.196$\pm$0.006&0.192$\pm$0.006&0.187$\pm$0.005&0.181$\pm$0.005&0.173$\pm$0.007&0.168$\pm$0.006&0.155$\pm$0.006 &0.135$\pm$0.007\\

\cline{2-12}
                   
                    & $R$ (fm) &5.823$\pm$0.009 & 3.384$\pm$0.006 & 1.402$\pm$0.005&1.118$\pm$0.006 &0.796$\pm$0.007&0.656$\pm$0.005&0.581$\pm$0.006&0.438$\pm$0.006&0.467$\pm$0.006&0.195$\pm$0.006\\
                   
                   \cline{2-12} 
                    & $\mu$  &-2.067$\pm$0.006 & -1.501$\pm$0.005&-0.647$\pm$0.005 &-0.498$\pm$0.008 &-0.123$\pm$0.006&0.040$\pm$0.006&0.140$\pm$0.005&0.351$\pm$0.005&0.439$\pm$0.005&0.785$\pm$0.009\\
                  \cline{2-12}
                    & $\chi^2$/ndf& 0.430&0.226&0.434&0.455&0.527&0.233&0.380&0.535&0.305&0.305\\
                      
\hline 
\multicolumn{2}{c|}{}&\multicolumn{2}{c|}{ Mul[1+2]}&\multicolumn{2}{c|}{ Mul[3+4]}&\multicolumn{2}{c|}{ Mul[5+6]}&\multicolumn{2}{c|}{ Mul[7+8]}&\multicolumn{2}{c|}{ Mul[9+10]}\\
\hline
\multirow{3}{*}{$\Omega^{-}+\bar{\Omega}^{+}$}& $T$ (GeV)  &\multicolumn{2}{c|}{0.207$\pm$0.007}&\multicolumn{2}{c|}{0.203$\pm$0.006}&\multicolumn{2}{c|}{0.192$\pm$0.006}&\multicolumn{2}{c|}{0.166$\pm$0.006}&\multicolumn{2}{c|}{0.138$\pm$0.007}\\

\cline{2-12} 
  	&  $R$ (fm)  &\multicolumn{2}{c|}{6.186$\pm$0.005} & \multicolumn{2}{c|}{1.637$\pm$0.007}&\multicolumn{2}{c|}{0.648$\pm$0.005}&\multicolumn{2}{c|}{0.413$\pm$0.006}&\multicolumn{2}{c|}{0.274$\pm$0.005} \\

 	\cline{2-12} 
  	& $\mu$ &\multicolumn{2}{c|}{-3.141$\pm$0.006} & \multicolumn{2}{c|}{-1.642$\pm$0.005}&\multicolumn{2}{c|}{-0.845$	\pm$0.006}&\multicolumn{2}{c|}{-0.306$\pm$0.008}&\multicolumn{2}{c|}{0.050$\pm$0.008} \\
 \cline{2-12}
 	& $\chi^2$/ndf& \multicolumn{2}{c|}{0.106}&\multicolumn{2}{c|}{1.624}&\multicolumn{2}{c|}{0.433}&\multicolumn{2}{c|}{1.524}&\multicolumn{2}{c|}{0.723}\\
\hline 
\end{tabular}
\end{adjustbox}
\end{table*}


\begin{figure*}[ht!]
\begin{center}
\includegraphics[scale = 0.44]{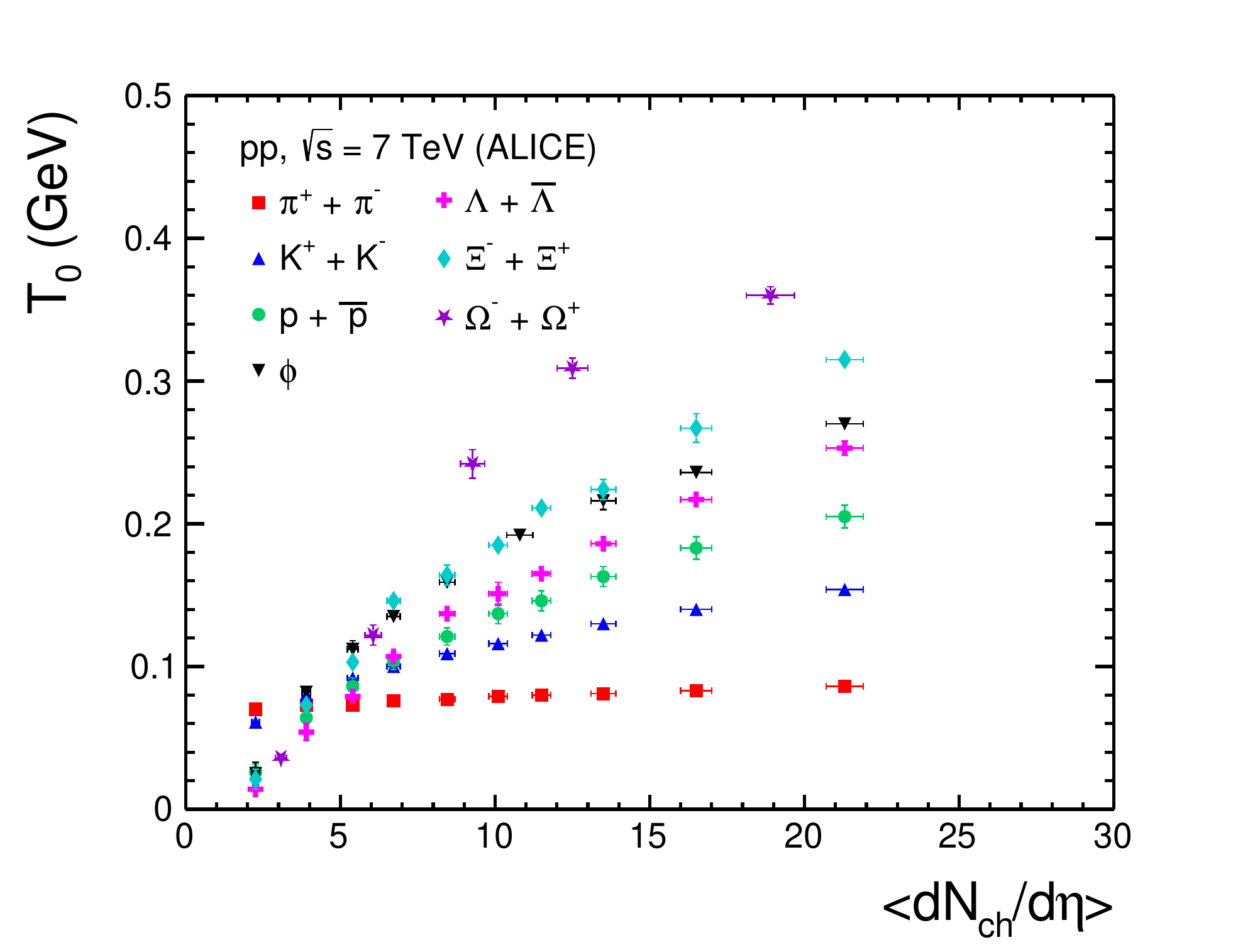}
\includegraphics[scale = 0.44]{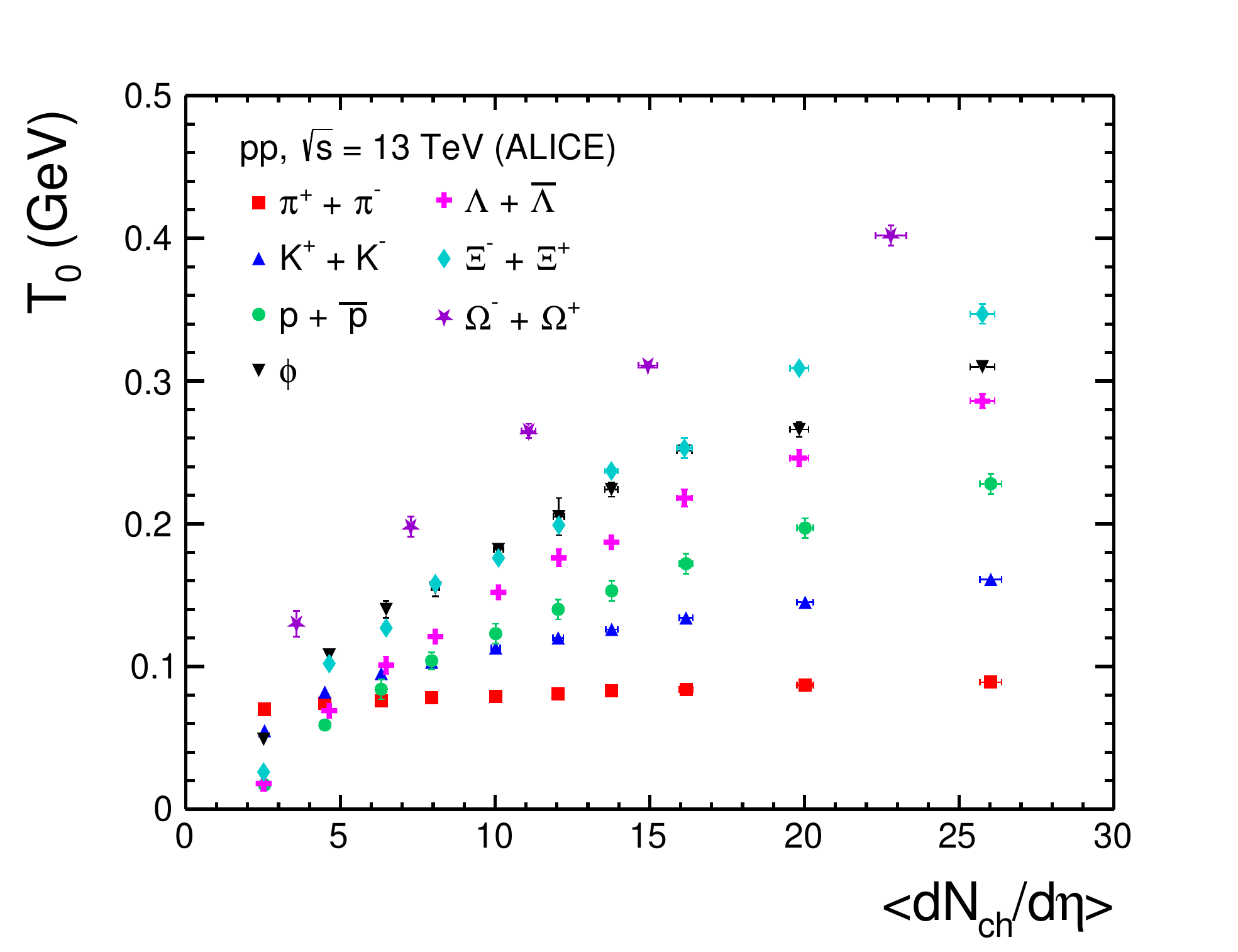}
\caption{(Color online) Temperature ($T_{0}$) as a function of charged-particle multiplicity for $pp$ collisions at $\sqrt{s}$ = 7 TeV (left panel) and 13 TeV (right panel) for different final state particles at zero chemical potential. The uncertainties in charged-particle multiplicity are the quadratic sum of statistical and systematic contributions, and the error in the value of $T_{0}$ are statistical errors.}
\label{fig4}
\end{center}
\end{figure*}

Further, Fig.~\ref{fig4} represents the temperature parameter, ($T_{0}$) extracted from the fitting as a function of charged-particle multiplicity for $pp$ collisions at $\sqrt{s}$ = 7 (left panel) and 13 TeV (right panel) for different final state particles at zero chemical potential. We observe that increasing charged-particle multiplicity increases the temperature for all the hadrons. A mass-ordering trend can be observed in the figures with heavier mass particles having a higher temperature than lighter mass particles at all charged-particle multiplicity. Similar results are obtained in Ref.~\cite{Khuntia:2018znt}. This corresponds to a mass-dependent differential freeze-out scenario, where particles freeze out at different times, corresponding to different volumes and temperatures for different particle species.

\begin{figure*}[ht!]
\begin{center}
\includegraphics[scale = 0.44]{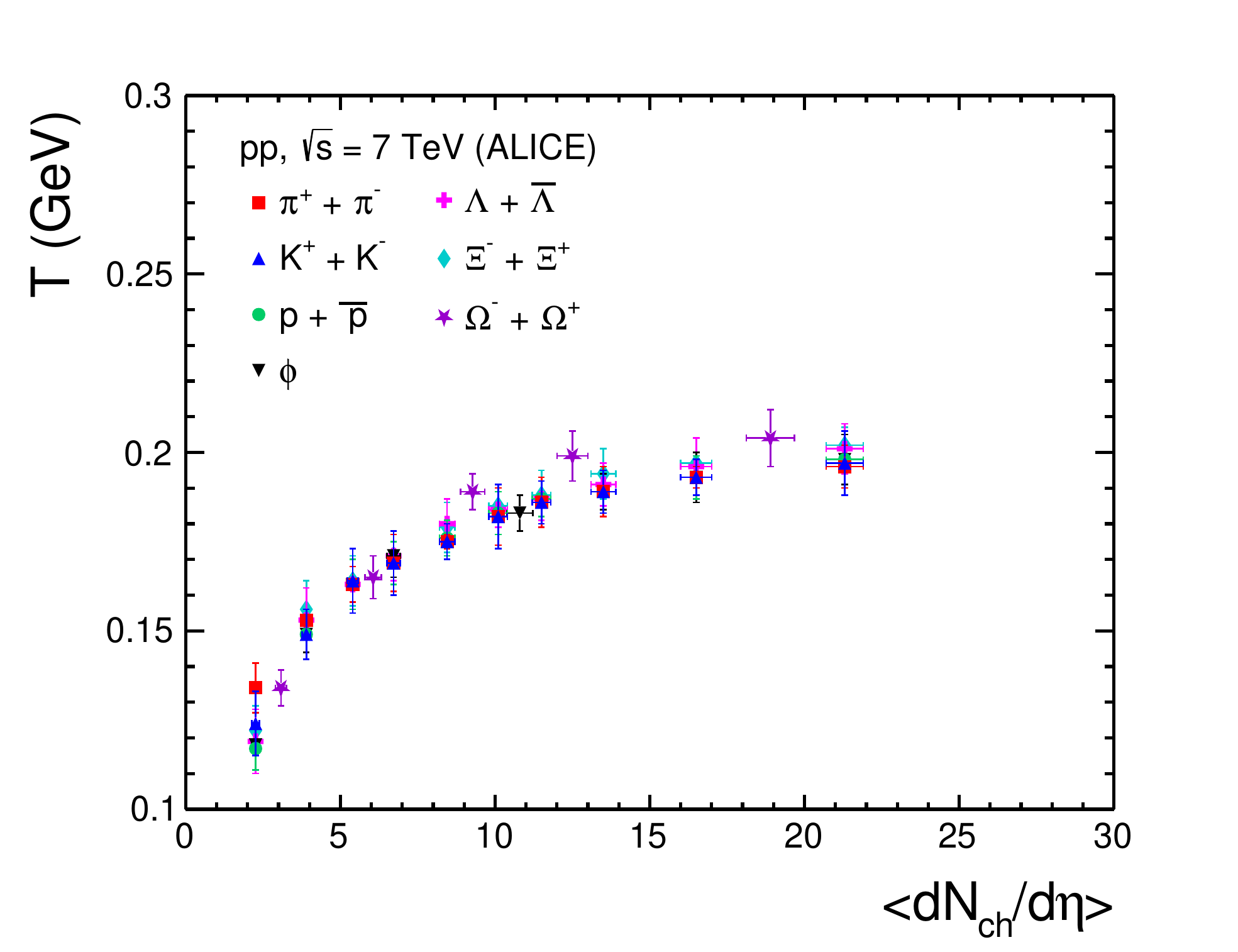}
\includegraphics[scale = 0.44]{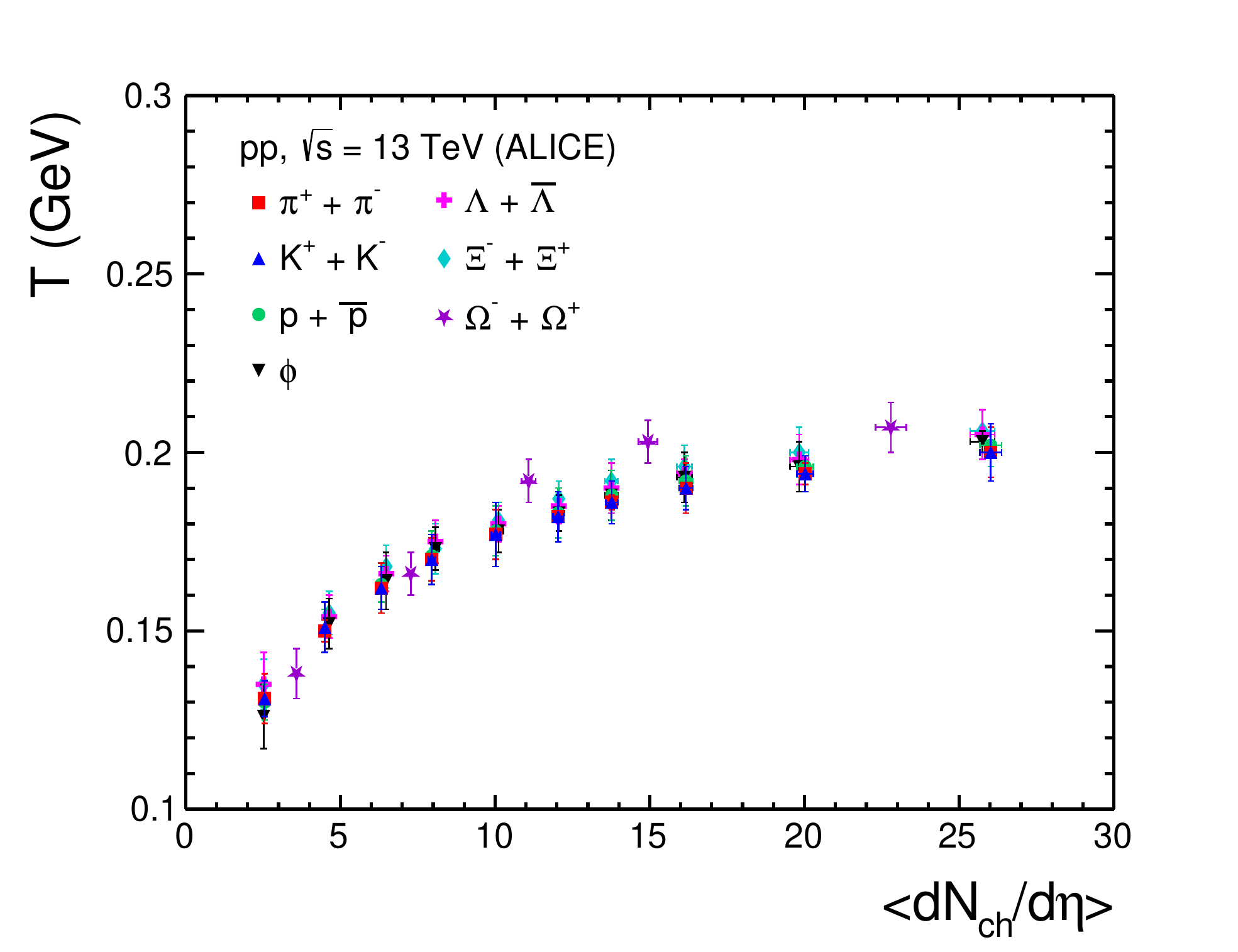}
\caption{(Color online) Temperature ($T$) as a function of charged-particle multiplicity for $pp$ collisions at $\sqrt{s}$ = 7 TeV (left panel) and 13 TeV (right panel) for different final state particles at the non-zero chemical potential. The uncertainties in charged-particle multiplicity are the quadratic sum of statistical and systematic contributions, and the error in the value of $T$ are statistical errors.}
\label{fig5}
\end{center}
\end{figure*}

Now, we use the $q$-values obtained from the first set of fits as fixed parameters for the second set of fits. In this case, the obtained parameters will change from $T_{0}, R_{0}$ to $T$ and $R$, and the chemical potential is kept as a free parameter. In Fig.~\ref{fig5}, we show the temperature ($T$) at non-zero chemical potential as a function of charged-particle multiplicity for $pp$ at $\sqrt{s}$ = 7  (left panel) and 13 TeV (right panel) for various final state particles. We observe a monotonic increase in the temperature for all the hadrons as we move towards the higher charged multiplicity. If we consider a given charged-particle multiplicity, the temperature shows a weak particle species dependency in $pp$ collision for both the center-of-mass energies. It suggests that all the particles have the same kinetic freeze-out temperature for a certain charged-particle multiplicity when we allow a finite $\mu$ for the system. At kinetic freeze-out, we observe a finite chemical potential. However, they do not have to be zero due to the absence of chemical equilibrium at kinetic freeze-out. 

\begin{figure*}[ht!]
\begin{center}
\includegraphics[scale = 0.44]{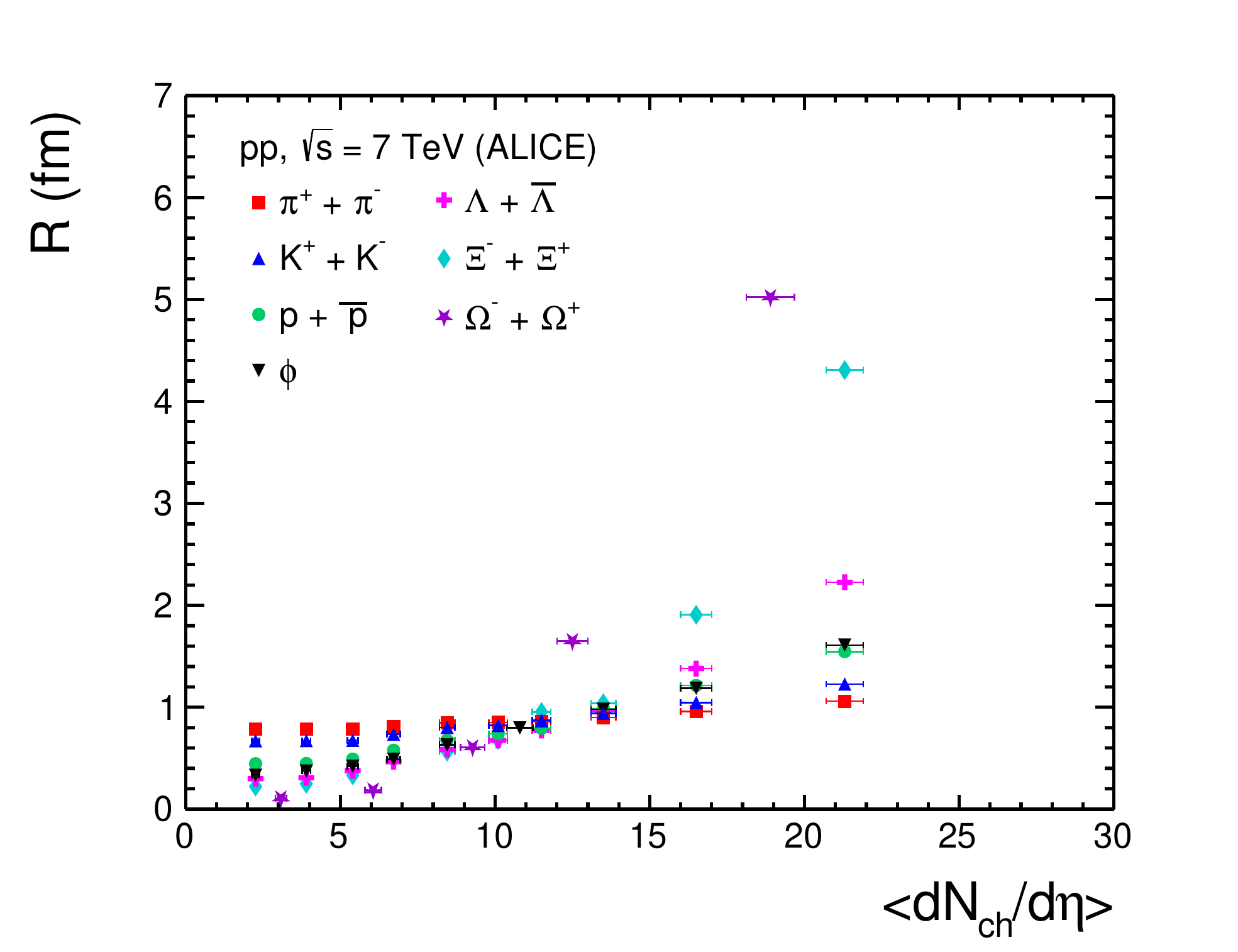}
\includegraphics[scale = 0.44]{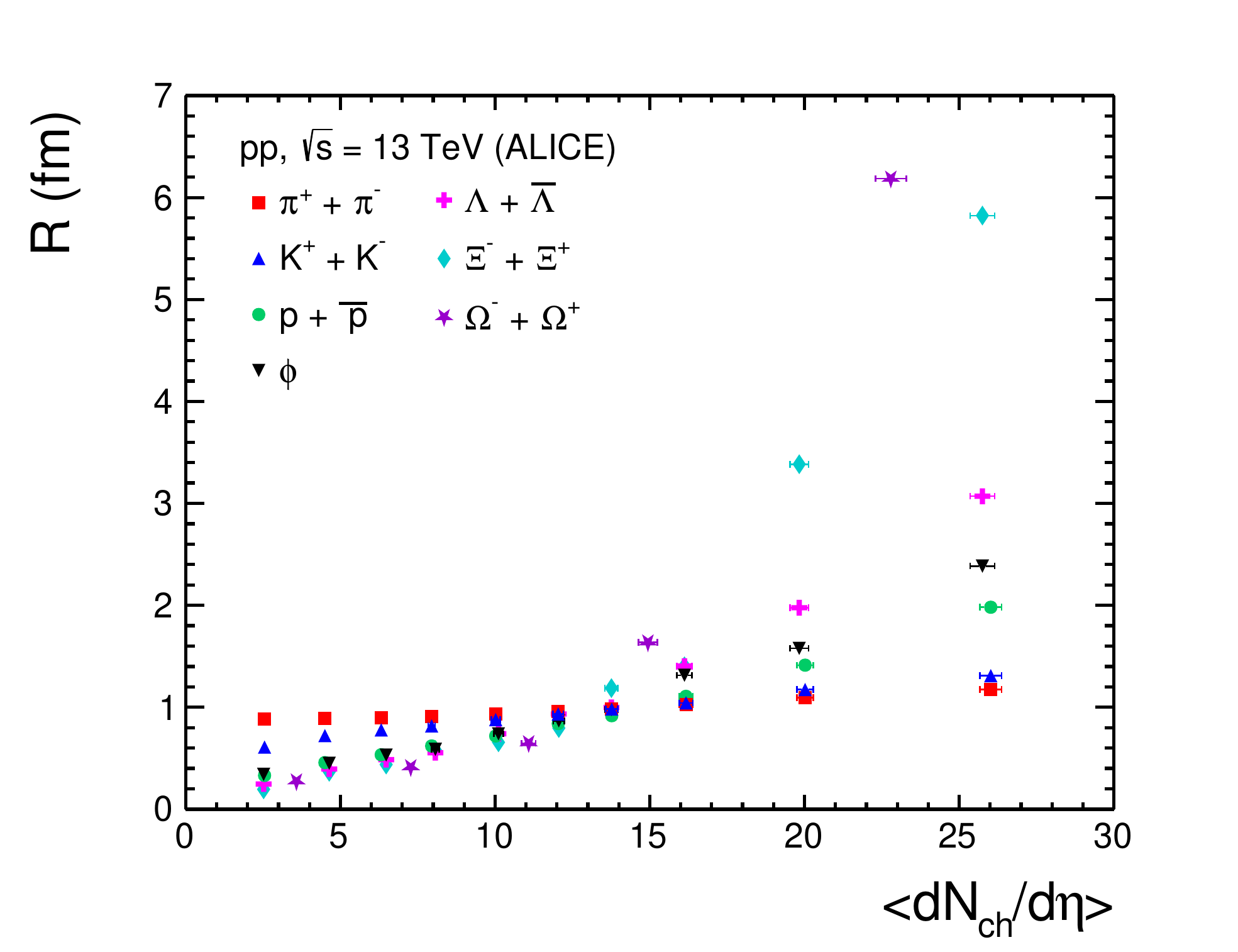}
\caption{(Color online) Radius of the system ($R$) as a function of charged-particle multiplicity for $pp$ collisions at $\sqrt{s}$ = 7 TeV (left panel) and 13 TeV (right panel) for different final state particles. The uncertainties in charged-particle multiplicity are the quadratic sum of statistical and systematic contributions, and the error in the value of $R$ are statistical errors.}
\label{fig6}
\end{center}
\end{figure*}

Figure~\ref{fig6} shows the radius of the system ($R$) as a function of charged-particle multiplicity for $pp$ collisions at $\sqrt{s}$ = 7 (left panel) and 13 TeV (right panel) for different final state particles at non-zero $\mu$. The value of $R$ increases with an increase in charged particle multiplicity, implying that as the number of charged particles produced in the collision increases, the size of the particle production region also increases. The increased charged particle multiplicity is often associated with more particle interactions and larger system sizes. We see a particle species dependency in both low and high multiplicity regions. However, we observe that the system's radius for all the hadrons taken in this study is almost identical between $\langle dN_{ch}/d\eta \rangle \simeq 8-14$. After that, we see a particle species dependency in the value when we move toward the higher charged-particle multiplicity. 

\begin{figure*}[ht!]
\begin{center}
\includegraphics[scale = 0.44]{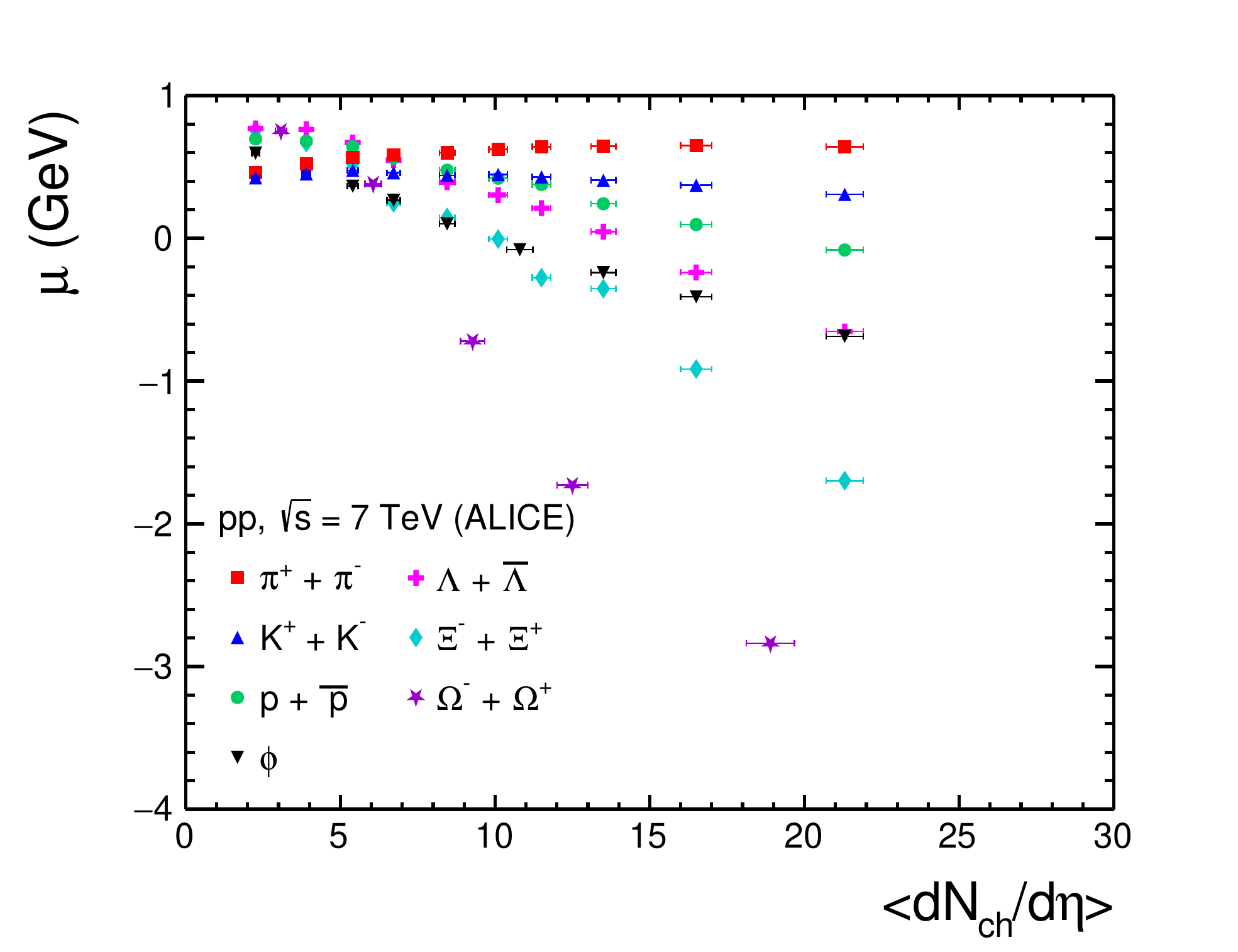}
\includegraphics[scale = 0.44]{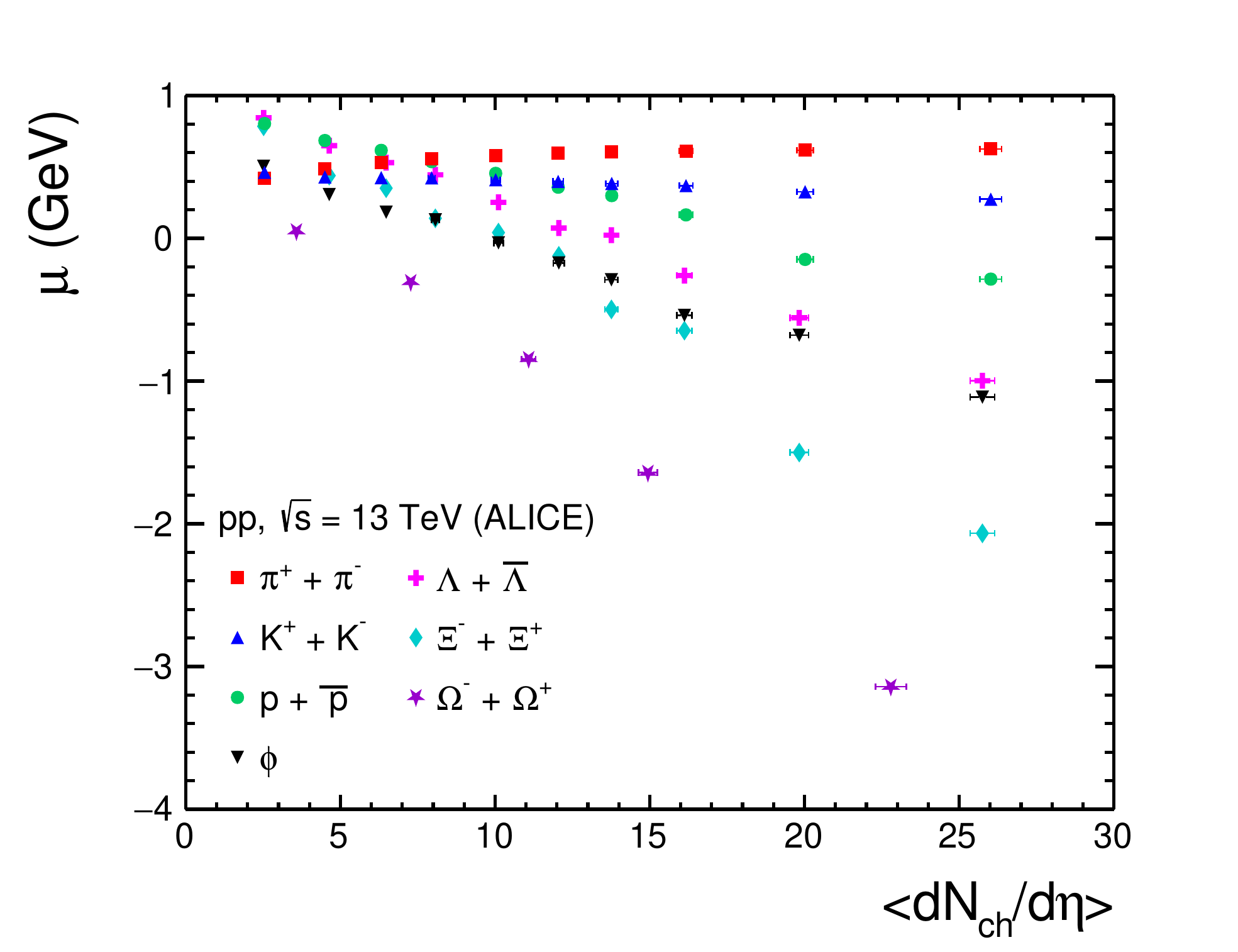}
\caption{(Color online) Chemical potential ($\mu$) as a function of charged-particle multiplicity for $pp$ collisions at $\sqrt{s}$ = 7 TeV (left panel) and 13 TeV (right panel) for different final state particles. The uncertainties in charged-particle multiplicity are the quadratic sum of statistical and systematic contributions, and the error in the value of $\mu$ are statistical errors.}
\label{fig7}
\end{center}
\end{figure*}

\begin{figure*}[ht!]
\begin{center}
\includegraphics[scale = 0.44]{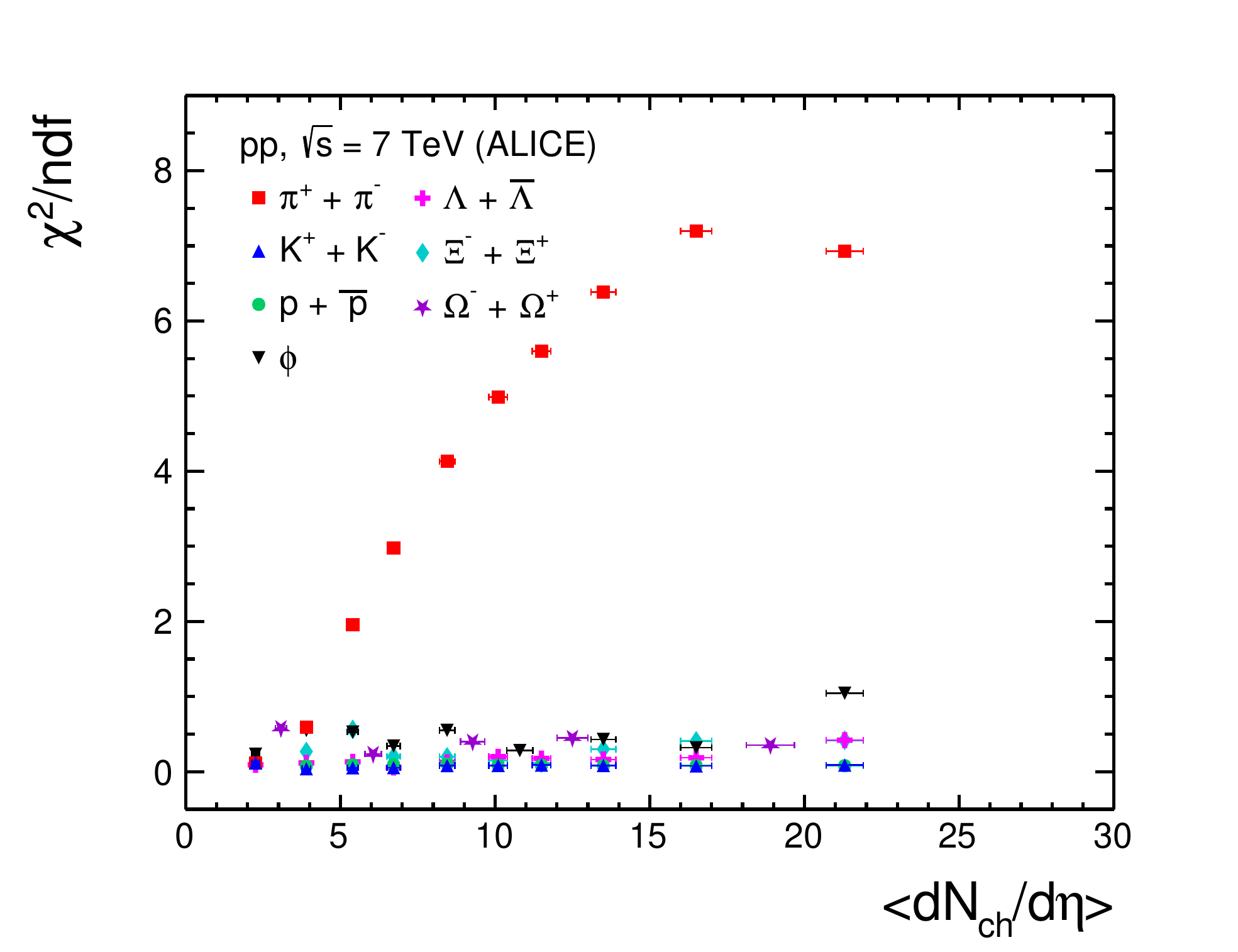}
\includegraphics[scale = 0.44]{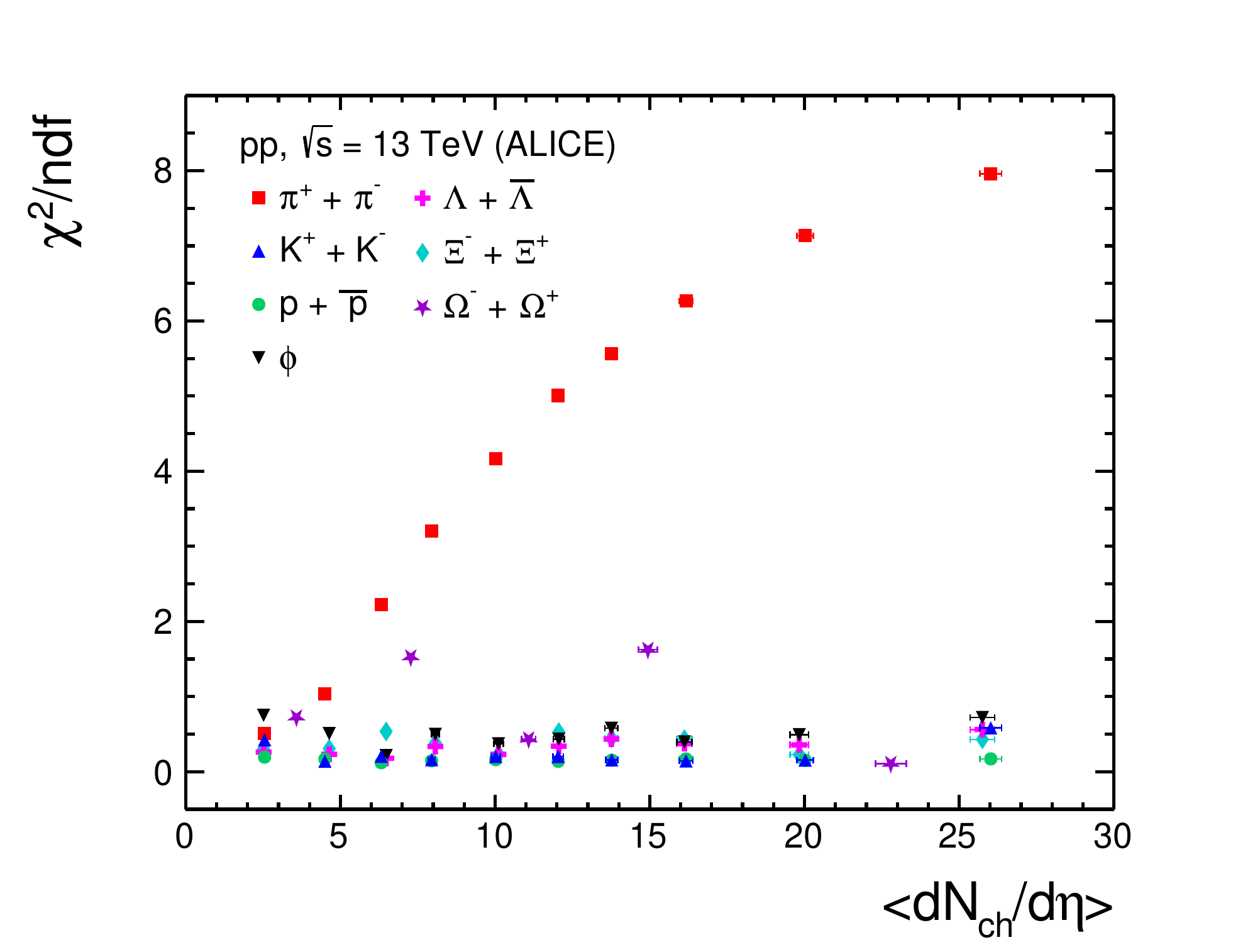}
\caption{(Color online) $\chi^{2}/ndf$ as a function of charged-particle multiplicity for $pp$ collisions at $\sqrt{s}$ = 7 TeV (left panel) and 13 TeV (right panel) for different final state particles.}
\label{fig8}
\end{center}
\end{figure*}

Figure~\ref{fig7} shows chemical potential ($\mu$) as a function of charged-particle multiplicity for $pp$ collisions at $\sqrt{s}$ = 7 (left panel) and 13 TeV (right panel) for different final state particles at a fixed value of non-extensive parameters. A non-zero value of the chemical potential at kinetic freeze-out temperature is observed for all the considered particle species. There is also a particle species dependency in the value of chemical potential for both the center-of-mass energies at the LHC. As we move towards massive particles, the chemical potentials become negative. However, we observe positive chemical potentials for all the charged-particle multiplicity for lighter particles such as $\pi$, $K$, and $p$. The $\mu$ values obtained for different particle types vary considerably. However, in the case of more massive particles, such as baryons, the chemical potential can transition from positive to negative as the rate of change of particle multiplicity with respect to pseudo rapidity increases. As we move towards higher massive particles, the chemical potential moves towards the negative value, which suggests that heavy mass particle production is less favorable.

We have taken the combined experimental spectra of particles and their antiparticle counterparts to obtain the chemical
potential as a function of the final state-charged particle multiplicity. Hence the observed value of the chemical potential is the absolute value. QGP-like signals are seen in the high-multiplicity class of $pp$ events at the LHC energies. To connect the hadron chemical potential to the constituent quarks, one assumes a quark-antiquark system at the ground state. Given
the chemical potential of a hadron, one can obtain the constituent quark chemical potentials by solving the following equation for a set of identified particles: $\mu^h = \sum_i \mu_i^q$, where $i$ runs from 1 to the number of constituent quarks of a hadron i.e. 2 for a meson and 3 for a baryon. Here $\mu^h$ is the chemical potential of a hadron and $\mu^q$
is the chemical potential of a quark/antiquark \cite{Bai:2019jtl}.

From the first law of thermodynamics, the chemical potential of a system is defined as the change in internal energy of the system when one more particle is added in such a way that the volume ($V$) and entropy ($S$) are constant, i.e. ($\mu = \frac{\partial U}{\partial N}|_{S, V}$ )\cite{Cook}. However, keeping entropy constant is a tricky part. For a system, the entropy increases with the increasing number of particles and also the increasing energy of the system. Therefore, one wants to increase the number of particles while keeping the entropy constant to define the system's chemical potential. For this, one must reduce the system's energy to compensate for the increasing effect on entropy as one increases the number of particles. At very high temperatures, the system of fermions and bosons behaves like a classical system, and they follow the classical distribution. To sum up, to increase the number of particles of a classical gas, one must change the system's energy by a negative amount. This change in energy is the chemical potential of a system. The chemical potential can take on different values depending on the system under consideration. In some cases, the chemical potential of a particle may be positive, indicating that adding more particles to the system requires energy input. On the other hand, a negative chemical potential implies that adding more particles to the system releases energy. 

In Fig.~\ref{fig8}, $\chi^{2}/ndf$ as a function of charged-particle multiplicity for $pp$ collisions at $\sqrt{s}$ = 7  (left panel) and 13 TeV (right panel) for different final state particles at non-zero chemical potential is shown. The quality of the fits is given by the reduced $\chi^{2}$. This indicates that the spectra are well described by the thermodynamically consistent form of Tsallis distribution. As the fit quality is very good up to 6 GeV in $p_{\rm T}$, the value of $\chi^{2}/ndf$ is always less than 1 except for pions. We also did the $p$-value test of the fitting and got $p$-value equal to 1 for all the cases.

\section{Summary}
\label{sum}

This article reviews another prospect to describe the kinetic freeze-out stage amidst significant chemical potential, specifically to explain the final state particles of the system produced in $pp$ collisions. A thorough analysis is presented, considering the chemical potential in the Tsallis distribution Eqn.~\ref{InvariantYield} by following a two-step procedure. We have used the redundancy present in the variables $T$, $V$, $q$, and $\mu$ expressed in Eqns.~\ref{T0} and \ref{V0} and performed all fit using Eqn.~\ref{YieldIndexZero}, that is effectively establishing the chemical potential equal to zero. This study reviewed a comparison of $T$ and $T_{0}$ values for both the center-of-mass energies. This result confirms that the variables $T$, $V$, $q$, and $\mu$ in the Tsallis distribution function Eqn.~\ref{InvariantYield} have a redundancy for $\mu \neq 0$.

The non-extensive Tsallis distribution describes the $p_{\rm T}$-spectra of identified hadrons very well in $pp$ collisions. However, a considerable $\chi^{2}/ndf$ value is obtained for pions. The resulting temperatures, $T$, are the same for all particle species considered in this study. In contrast to the results obtained in Ref.~\cite{Bhattacharyya:2017hdc}, the inconsistency can be traced back to the incorrect use of chemical potentials in the last analysis.
The kinetic freeze-out temperature for all the particle species studied in this analysis seems to be the same when one considers a finite chemical potential at the kinetic freeze-out of the produced fireball. The $\mu$ values obtained for different particle species vary considerably, and there is no chemical equilibrium at kinetic freeze-out in $pp$ collisions. This work explores the importance of the chemical potential in the Tsallis distribution while analyzing the identified particle spectra in $pp$ collisions. This leads to a detailed analysis of various parameters in the new domain and confirms the usefulness of the Tsallis distribution in high-energy collisions.

\section*{Acknowledgement} 
GSP acknowledges the financial support from the DST-INSPIRE program of the Government of India. DS and RS gratefully acknowledge the DAE-DST, Govt. of India funding under the mega-science project – “Indian participation in the ALICE experiment at CERN” bearing Project No. SR/MF/PS-02/2021-IITI (E-37123). RR acknowledges the support under the INFN postdoctoral fellowship.

{}


\begin{thebibliography}{}

\bibitem{Gyulassy:2004zy}
M.~Gyulassy and L.~McLerran,
Nucl. Phys. A \textbf{750}, 30 (2005)

\bibitem{Braun-Munzinger:2007edi}
P.~Braun-Munzinger and J.~Stachel,
Nature \textbf{448}, 302 (2007)

\bibitem{Jacak:2012dx}
B.~V.~Jacak and B.~Muller,
Science \textbf{337}, 310 (2012)

\bibitem{Itoh:1970uw}
N.~Itoh,
Prog. Theor. Phys. \textbf{44}, 291 (1970)

\bibitem{Collins:1974ky}
J.~C.~Collins and M.~J.~Perry,
Phys. Rev. Lett. \textbf{34}, 1353 (1975)

\bibitem{Cabibbo:1975ig}
N.~Cabibbo and G.~Parisi,
Phys. Lett. B \textbf{59}, 67 (1975)

\bibitem{Chapline:1976gy}
G.~Chapline and M.~Nauenberg,
Phys. Rev. D \textbf{16}, 450 (1977)

\bibitem{Waqas:2019bnc}
M.~Waqas and F.~H.~Liu,
Eur. Phys. J. Plus \textbf{135}, 147 (2020)

\bibitem{Wang:2021ccs}
Q.~Wang, F.~H.~Liu and K.~K.~Olimov,
Front. in Phys. \textbf{9}, 792039 (2021)

\bibitem{Cleymans:1999st}
J.~Cleymans and K.~Redlich,
Phys. Rev. C \textbf{60}, 054908 (1999)

\bibitem{Florkowski:1999pz}
W.~Florkowski and W.~Broniowski,
Phys. Lett. B \textbf{477}, 73 (2000)

\bibitem{Braun-Munzinger:2003htr}
P.~Braun-Munzinger, J.~Stachel and C.~Wetterich,
Phys. Lett. B \textbf{596}, 61 (2004)

\bibitem{Waqas:2018xrz}
M.~Waqas, F.~H.~Liu, S.~Fakhraddin and M.~A.~Rahim,
Indian J. Phys. \textbf{93}, 1329 (2019)

\bibitem{Waqas:2021bsy}
M.~Waqas and G.~X.~Peng,
Entropy \textbf{23}, 488 (2021)

\bibitem{Waqas:2021enm}
M.~Waqas, H.~M.~Chen, G.~X.~Peng, A.~A.~K.~H.~Ismail, M.~Ajaz, Z.~Wazir, R.~Shehzadi, S.~Jamal and A.~AbdelKader,
Entropy \textbf{23}, 1363 (2021)

\bibitem{Ajaz:2021awb}
M.~Ajaz, M.~Waqas, G.~X.~Peng, Z.~Yasin, H.~Younis and A.~A.~K.~H.~I.~l,
Eur. Phys. J. Plus \textbf{137}, 52 (2022)

\bibitem{Andronic:2017pug}
A.~Andronic, P.~Braun-Munzinger, K.~Redlich and J.~Stachel,
Nature \textbf{561}, 321 (2018)

\bibitem{Sollfrank:1990qz}
J.~Sollfrank, P.~Koch and U.~W.~Heinz,
Phys. Lett. B \textbf{252}, 256 (1990)

\bibitem{Tang:2008ud}
Z.~Tang, Y.~Xu, L.~Ruan, G.~van Buren, F.~Wang and Z.~Xu,
Phys. Rev. C \textbf{79}, 051901 (2009)

\bibitem{Chatterjee:2014lfa}
S.~Chatterjee, B.~Mohanty and R.~Singh,
Phys. Rev. C \textbf{92}, 024917 (2015)

\bibitem{Chatterjee:2015fua}
S.~Chatterjee, S.~Das, L.~Kumar, D.~Mishra, B.~Mohanty, R.~Sahoo and N.~Sharma,
Adv. High Energy Phys. \textbf{2015}, 349013 (2015)


\bibitem{Thakur:2016boy}
D.~Thakur, S.~Tripathy, P.~Garg, R.~Sahoo and J.~Cleymans,
Adv. High Energy Phys. \textbf{2016}, 4149352 (2016)

\bibitem{Chatterjee:2014ysa}
S.~Chatterjee and B.~Mohanty,
Phys. Rev. C \textbf{90}, 034908 (2014)

\bibitem{Waqas:2018tkk}
M.~Waqas and F.~H.~Liu,
Indian J. Phys. \textbf{96}, 1217 (2022)

\bibitem{Waqas:2021jph}
M.~Waqas, G.~X.~Peng, F.~H.~Liu and Z.~Wazir,
Sci. Rep. \textbf{11}, 20252 (2021)

\bibitem{Waqas:2021rmb}
M.~Waqas, G.~X.~Peng and F.~H.~Liu,
J. Phys. G \textbf{48}, 075108 (2021)

\bibitem{ALICE:2020nkc}
S.~Acharya \textit{et al.} [ALICE],
Eur. Phys. J. C \textbf{80}, 693 (2020)

\bibitem{CMS:2015fgy}
V.~Khachatryan \textit{et al.} [CMS],
Phys. Rev. Lett. \textbf{116}, 172302 (2016)

\bibitem{Allton:2002zi}
C.~R.~Allton, S.~Ejiri, S.~J.~Hands, O.~Kaczmarek, F.~Karsch, E.~Laermann, C.~Schmidt and L.~Scorzato,
Phys. Rev. D \textbf{66}, 074507 (2002)

\bibitem{Allton:2003vx}
C.~R.~Allton, S.~Ejiri, S.~J.~Hands, O.~Kaczmarek, F.~Karsch, E.~Laermann and C.~Schmidt,
Phys. Rev. D \textbf{68}, 014507 (2003)

\bibitem{Gavai:2003mf}
R.~V.~Gavai and S.~Gupta,
Phys. Rev. D \textbf{68}, 034506 (2003)


\bibitem{Allton:2005gk}
C.~R.~Allton, M.~Doring, S.~Ejiri, S.~J.~Hands, O.~Kaczmarek, F.~Karsch, E.~Laermann and K.~Redlich,
Phys. Rev. D \textbf{71}, 054508 (2005)

\bibitem{Bazavov:2017dus}
A.~Bazavov, H.~T.~Ding, P.~Hegde, O.~Kaczmarek, F.~Karsch, E.~Laermann, Y.~Maezawa, S.~Mukherjee, H.~Ohno and P.~Petreczky, \textit{et al.}
Phys. Rev. D \textbf{95}, 054504 (2017)


\bibitem{Book} 
K.~Rajagopal and F.~Wilczek, in {\sl At the Frontier of Particle
Physics: Handbook of QCD\/} 
ed. M. Shifman, p. 2061 (World Scientific, Singapore) (2001)

\bibitem{Abelev:2006cs}
B.~I.~Abelev \textit{et al.} [STAR Collaboration],
Phys. Rev. C \textbf{75}, 064901 (2007)

\bibitem{PHENIX:2010qqf}
A.~Adare \textit{et al.} [PHENIX],
Phys. Rev. D \textbf{83}, 052004 (2011)

\bibitem{PHENIX:2011rvu}
A.~Adare \textit{et al.} [PHENIX],
Phys. Rev. C \textbf{83}, 064903 (2011)

\bibitem{CMS:2010wcx}
V.~Khachatryan \textit{et al.} [CMS],
JHEP \textbf{02}, 041 (2010)

\bibitem{CMS:2010tjh}
V.~Khachatryan \textit{et al.} [CMS],
Phys. Rev. Lett. \textbf{105}, 022002 (2010)

\bibitem{ATLAS:2010jvh}
G.~Aad \textit{et al.} [ATLAS],
New J. Phys. \textbf{13}, 053033 (2011)

\bibitem{Aamodt:2011zj}
K.~Aamodt \textit{et al.} [ALICE Collaboration],
Eur. Phys. J. C \textbf{71}, 1655 (2011)

\bibitem{Abelev:2012cn}
B.~Abelev \textit{et al.} [ALICE Collaboration],
Phys. Lett. B \textbf{717}, 162 (2012)

\bibitem{Abelev:2012jp}
B.~Abelev \textit{et al.} [ALICE Collaboration],
Phys. Lett. B \textbf{712}, 309 (2012)

\bibitem{Chatrchyan:2012qb}
S.~Chatrchyan \textit{et al.} [CMS Collaboration],
Eur. Phys. J. C \textbf{72}, 2164 (2012)


\bibitem{Cleymans:2011in}
J.~Cleymans and D.~Worku,
J. Phys. G \textbf{39}, 025006 (2012)

\bibitem{Cleymans:2012ya}
J.~Cleymans and D.~Worku,
Eur. Phys. J. A \textbf{48}, 160 (2012)

\bibitem{Azmi:2015xqa}
M.~D.~Azmi and J.~Cleymans,
Eur. Phys. J. C \textbf{75}, 430 (2015)

\bibitem{Pradhan:2021vtp}
G.~S.~Pradhan, D.~Sahu, S.~Deb and R.~Sahoo,
J. Phys. G \textbf{50}, 055104 (2023)

\bibitem{Tsallis:1998ws}
C.~Tsallis, R.~S.~Mendes and A.~R.~Plastino,
Physica A \textbf{261}, 534 (1998)

\bibitem{Tsallis:1987eu}
C.~Tsallis,
J. Statist. Phys. \textbf{52}, 479 (1988)

\bibitem{Lyra:1997ggy}
M.~L.~Lyra and C.~Tsallis,
Phys. Rev. Lett. \textbf{80}, 53 (1998)

\bibitem{Tsallis:2012js}
C.~Tsallis and L.~J.~L.~Cirto,
Eur. Phys. J. C \textbf{73}, 2487 (2013)

\bibitem{Luciano:2021ndh}
G.~G.~Luciano,
Eur. Phys. J. C \textbf{81}, 672 (2021)

\bibitem{Tsallis:book}
C.~Tsallis, Introduction to Non-Extensive Statistical Mechanics: Approaching a Complex World (Springer, Berlin, 2009) 

\bibitem{Plastino}
Plastino, A.R. and Plastino, A. 
Phys. Lett. A, 174, 384 (1993)

\bibitem{Tsallis:1995zza}
C.~Tsallis, F.~C.~Sa Barreto and E.~D.~Loh,
Phys. Rev. E \textbf{52}, 1447 (1995)

\bibitem{Jizba}
P.~Jizba, J.~Korbel, V.~Zatloukal, 
Phys. Rev. E 95, 022103 (2017) 

\bibitem{Luciano:2021mto}
G.~G.~Luciano and M.~Blasone,
Phys. Rev. D \textbf{104}, 045004 (2021)

\bibitem{Parvan:2021rhi}
A.~S.~Parvan,
Physica A: Statistical Mechanics and its Applications \textbf{588}, 126556 (2022)

\bibitem{Parvan:2019aii}
A.~S.~Parvan and T.~Bhattacharyya,
Eur. Phys. J. A \textbf{56}, 72 (2020)

\bibitem{Wilk:1999dr}
G.~Wilk and Z.~Wlodarczyk,
Phys. Rev. Lett. \textbf{84}, 2770 (2000)

\bibitem{Biro:2004qg}
T.~S.~Biro and A.~Jakovac,
Phys. Rev. Lett. \textbf{94}, 132302 (2005)

\bibitem{Wilk:2009nn}
G.~Wilk and Z.~Wlodarczyk,
Phys. Rev. C \textbf{79}, 054903 (2009)

\bibitem{Wilk:2008ue}
G.~Wilk and Z.~Wlodarczyk,
Eur. Phys. J. A \textbf{40}, 299 (2009)

\bibitem{Biro:2014yoa}
T.~S.~Biro, G.~G.~Barnaf\"oldi and P.~Van,
Physica A \textbf{417}, 215 (2015)


\bibitem{Bhattacharyya:2015nwa}
T.~Bhattacharyya, P.~Garg, R.~Sahoo and P.~Samantray,
Eur. Phys. J. A \textbf{52}, 283 (2016).

\bibitem{Deppman:2012qt}
A.~Deppman,
J. Phys. G \textbf{41}, 055108 (2014)

\bibitem{Deppman:2016fxs}
A.~Deppman,
Phys. Rev. D \textbf{93}, 054001 (2016)

\bibitem{Deppman:2019yno}
A.~Deppman, E.~Megias and D.~P.~Menezes,
Phys. Rev. D \textbf{101}, no.3, 034019 (2020)

\bibitem{Deppman:2017fkq}
A.~Deppman, E.~Megias, D.~P.~Menezes and T.~Frederico,
Entropy \textbf{20}, 633 (2018)


\bibitem{Cleymans:2013rfq}
J.~Cleymans, G.~I.~Lykasov, A.~S.~Parvan, A.~S.~Sorin, O.~V.~Teryaev and D.~Worku,
Phys. Lett. B \textbf{723}, 351 (2013)

\bibitem{Cleymans:2020ojr}
J.~Cleymans and M.~Wellington Paradza,
MDPI Physics \textbf{2}, 654 (2020)

\bibitem{Cleymans:2020nvs}
J.~Cleymans and M.~W.~Paradza,
[arXiv:2010.05565 [hep-ph]]

\bibitem{Rybczynski:2014cha}
M.~Rybczynski and Z.~Wlodarczyk,
Eur. Phys. J. C \textbf{74}, 2785 (2014)

\bibitem{Okorokov:2014cna}
V.~A.~Okorokov,
Adv. High Energy Phys. \textbf{2015}, 790646 (2015)

\bibitem{Okorokov:2016vug}
V.~A.~Okorokov,
Adv. High Energy Phys. \textbf{2016}, 5972709 (2016)

\bibitem{ALICE:2010igk}
K.~Aamodt \textit{et al.} [ALICE],
Phys. Rev. D \textbf{82}, 052001 (2010)

\bibitem{STAR:2010yvd}
M.~M.~Aggarwal \textit{et al.} [STAR],
Phys. Rev. C \textbf{83}, 064905 (2011)



\bibitem{ALICE:2011kmy}
K.~Aamodt \textit{et al.} [ALICE],
Phys. Rev. D \textbf{84}, 112004 (2011)

\bibitem{ROOT}
Version: 6.26/06, CERN ROOT: http://root.cern.ch.


\bibitem{Khuntia:2018znt}
A.~Khuntia, H.~Sharma, S.~Kumar Tiwari, R.~Sahoo and J.~Cleymans,
Eur. Phys. J. A \textbf{55}, 3 (2019)


\bibitem{ALICE:2018pal}
S.~Acharya \textit{et al.} [ALICE],
Phys. Rev. C \textbf{99}, 024906 (2019)

\bibitem{ALICE:2019etb}
S.~Acharya \textit{et al.} [ALICE],
Phys. Lett. B \textbf{807}, 135501 (2020)

\bibitem{ALICE:2019avo}
S.~Acharya \textit{et al.} [ALICE],
Eur. Phys. J. C \textbf{80}, 167 (2020)

\bibitem{Bai:2019jtl}
Z.~Bai and Y.~X.~Liu,
Phys. Rev. D \textbf{108}, 014018 (2023)

\bibitem{Cook}
G. Cook, R. H. Dickerson, 
American Journal of Physics 63, 737 (1995)

\bibitem{Bhattacharyya:2017hdc}
T.~Bhattacharyya, J.~Cleymans, L.~Marques, S.~Mogliacci, and M.~W.~Paradza,
J. Phys. G \textbf{45}, 055001 (2018)


\end{thebibliography}
\end{document}